\documentclass[11pt,a4paper]{article}

\usepackage[english]{babel}

\usepackage{amsmath}
\usepackage{a4,amssymb,amstext,theorem,enumerate}
\usepackage{amsfonts}
\usepackage{latexsym,epic,graphicx}

\usepackage{fancyhdr}

\makeatletter
\renewcommand{\theequation}{\thesection.\arabic{equation}}
\@addtoreset{equation}{section}
\makeatother

\newcommand{\be}{\begin{eqnarray}}
\newcommand{\ee}{\end{eqnarray}}
\newcommand{\beg}{\begin{eqnarray*}}
\newcommand{\eeg}{\end{eqnarray*}}

\newcommand{\nn}{\nonumber}
\newcommand{\Spin}{\textit{Spin}}
\newcommand{\Diff}{\textit{Diff}}

\newcommand{\e}{\epsilon}
\newcommand{\ep}{\varepsilon}
\newcommand{\eps}{\epsilon}

\newcommand{\p}{\partial}
\newcommand{\G}{\Gamma}
\newcommand{\g}{\gamma}

\newcommand{\cB}{\mathcal{B}}
\newcommand{\cE}{\mathcal{E}}
\newcommand{\cEs}{\mathcal{E}^\#}

\newcommand{\cF}{\mathcal{F}}
\newcommand{\cL}{\mathcal{L}}
\newcommand{\cO}{\mathcal{O}}

\newcommand{\cA}{\mathcal{A}}
\newcommand{\cV}{\mathcal{V}}
\newcommand{\deltaS}{\underline{\delta}}
\newcommand{\cN}{\mathcal{N}}

\newcommand{\cSU}{\Phi}
\newcommand{\cpiP}{\tilde{\pi}}

\newcommand{\cH}{\mathcal{H}}
\newcommand{\cpi}{\pi}
\newcommand{\cpig}{\pi}
\newcommand{\cpiV}{\pi}
\newcommand{\cPhi}{\Phi}

\newcommand{\Z}{\mathbb{Z}}
\newcommand{\R}{\mathbb{R}}

\newcommand{\cW}{W}
\newcommand{\cvB}{\mathcal{Q}} 
\newcommand{\cvA}{\mathcal{P}} 
\newcommand{\cMa}{\texttt{m}} 
\newcommand{\cNa}{\texttt{n}} 
\newcommand{\cPa}{\texttt{p}}
\newcommand{\cQa}{\texttt{q}} 
\newcommand{\h}{h}

\newcommand{\Rea}{\operatorname{Re}}
\newcommand{\Ima}{\operatorname{Im}}

\newcommand{\id}{\mathbf{1\hspace{-2.9pt}l}}

\begin{document}

{\flushright IHES/P/09/52\\[9mm]}

\renewcommand{\thefootnote}{\fnsymbol{footnote}}

\begin{center}
{\LARGE \bf $E_{7(7)}$ invariant Lagrangian of $d=4$ $\cN=8$ supergravity  }\\[1cm]
Christian Hillmann\footnote[3]{E-mail: \tt hillmann@ihes.fr} 
\\[5mm]
{\sl  Institut des Hautes \'Etudes Scientifiques\\
           35, Route de Chartres\\
      91440 Bures-sur-Yvette, France} \\[15mm]

\begin{tabular}{p{12cm}}
\hline\\\hspace{5mm}{\bf Abstract:}
We present an $E_{7(7)}$ invariant Lagrangian that leads to the equations of motion of $d=4$ $\cN=8$ supergravity without using Lagrange multipliers. The superinvariance of this new action and the closure of the supersymmetry algebra are proved explicitly for the terms that differ from the Cremmer--Julia formulation. Since the diffeomorphism symmetry is not realized in the standard way on the vector fields, we switch to the Hamiltonian formulation in order to prove the invariance of the $E_{7(7)}$ invariant action under general coordinate transformations. We also construct the conserved $E_{7(7)}$-Noether current of maximal supergravity and we conclude with comments on the implications of this manifest off-shell $E_{7(7)}$-symmetry for quantizing $d=4$ $\cN=8$ supergravity, in particular on the $E_{7(7)}$-action on phase space.
\\\\\hline
\end{tabular}\\[9mm]
\end{center}

\renewcommand{\thefootnote}{\arabic{footnote}}
\setcounter{footnote}{0}

\begin{section}{Introduction}
The appearance of the hidden $E_{7(7)}$ symmetry is one of the most remarkable features of maximal supergravity in four dimensions \cite{CJ79}, but its origin is still quite mysterious. In the standard formulation \cite{CJ79,dWN82}, $E_{7(7)}$ only is a symmetry of the equations of motion, but it is broken on the level of the Lagrangian. This statement is based on the well-known fact that $D=11$ supergravity \cite{CJS78} only gives rise to $28$ vector fields upon a reduction \`a la Kaluza--Klein on a flat seven torus to $d=4$. Since the smallest, non-trivial $E_{7(7)}$ representation has dimension $56$, $E_{7(7)}$ only becomes a manifest symmetry of the equations of motion upon combining the field equations of the vector fields and their Bianchi identities into one equation \cite{CJ79,dWN82}. Another possibility is to start with a manifestly $E_{7(7)}$-invariant Lagrangian containing $56$ vector fields and to impose a twisted self-duality constraint on the vector fields on top of the equations of motion \cite{CJLP97}. \\

In this article, we shall prove that there exists an off-shell formulation of $d=4$ $\cN=8$ supergravity that exhibits $E_{7(7)}$-symmetry manifestly without imposing a constraint nor using Lagrange multipliers.\footnote{Note that an $E_{7(7)}$ invariant Lagrangian with a Lagrange multiplier has been stated in eq. (6.27) of \cite{CJ79}, which leads to the well-known problems upon quantization \cite{HT88}. What is more, the superinvariance of this formulation has not been shown.} Our approach differs in so far as that we start with a manifestly $E_{7(7)}$ invariant Lagrangian (that contains in particular $56$ vector fields). The associated equations of motion are shown to exactly coincide with the ones of the standard formulation of maximal supergravity without imposing any further constraint. The price to pay for this is that we have to dispense with the usual form of manifest four-dimensional general coordinate covariance on the level of the action, because we shall adopt an ADM-split into time and space \cite{ADM62}. Nevertheless, the action will be proved to exhibit invariance under general coordinate transformations closely following the arguments pioneered by Henneaux and Teitelboim in \cite{HT88}. Hence, it does not come as a surprise that all the equations of motion do in fact exhibit both $E_{7(7)}$- and $\Diff(4)$-covariance explicitly.\\

This paper is structured as follows: We start by discussing the part of the bosonic Lagrangian containing the vector fields alone, before coupling it to the other bosonic terms in the action. In order to couple consistently to the fermions of supergravity, we shall then switch to the flat ``vielbein frame'' for both the coordinate indices and the $E_{7(7)}$-indices, where we use the scalars of maximal supergravity as a ``vielbein'', i.e. to intertwine between the $E_{7(7)}$-covariant and the $SU(8)$-covariant formulation in the standard way \cite{CJ79}. As a next step, we add the fermionic part of the action, which will naturally lead to the supercovariant extension of the vector field strengths. We verify the closure of the supersymmetry algebra on the bosons as well as the superinvariance of the action functional in our manifestly $E_{7(7)}$-invariant formulation. As a next step, we extract the conserved $E_{7(7)}$-Noether current, before switching to the Hamiltonian formulation of the theory. This allows to prove the invariance of the action under general coordinate transformations and furthermore reveals that the Noether charge of the duality symmetry $E_{7(7)}$ shows exactly the same properties as the one of an ordinary global symmetry. We conclude with a computation of the relevant Dirac brackets and with an analysis of the phase space of maximal supergravity from the $E_{7(7)}$-symmetric point of view that is expected to improve our understanding of the quantization of maximal supergravity in four dimensions.
\end{section}

\begin{section}{Bosonic dynamics}
The usual argument against an off-shell $E_{7(7)}$-symmetry in maximal supergravity is related to the counting of the degrees of freedom. The important observation is however that the number of the on-shell degrees of freedom is intimately linked to the form of the equations of motion. For {\textit{second-order}} Maxwell-type ones, i.e. $dF=0$ and $d*F=0$, it would clearly be inconsistent to keep $56$ vector fields forming the lowest dimensional, non-trivial representation of $E_{7(7)}$, because it would violate the equality of bosonic and fermionic degrees of freedom in the theory, which restricts the number of the vector fields to $28$ \cite{CJ79}. The key idea in this paper is that we are looking for an action involving $56$ vector fields that gives rise to a different set of equations of motion directly, namely {\textit{first-order}} twisted self-duality equations of motion \cite{CJLP97} that exhibit $\Diff(4)$- and $E_{7(7)}$-invariance at the same time. Thus, the counting of degrees of freedom will match again and it will in particular not be necessary to impose a twisted self-duality constraint on top of the equations of motion as was done in \cite{CJLP97}.\\

For a better readability of the article, we have separated the fermionic part $S_{\text{ferm}}$ from the complete action $S=S_{\text{bos}} +S_{\text{ferm}}$ and we have divided the bosonic part $S_{\text{bos}}$ into three pieces
\beg
S_{\text{bos}}&=& S_{\text{grav}} +S_{\text{scal}}+S_{\text{vec}}.
\eeg
The first term $S_{\text{grav}}$ is the usual Einstein--Hilbert action in four dimensions, to which the scalars are coupled by the standard $\sigma$-model action $S_{\text{scal}}$. The last term $S_{\text{vec}}$ in the bosonic part of the action describes the dynamics of the $56$ vector fields and their coupling to the metric and to the scalars. We will start by stating $S_{\text{vec}}$ and by proving that the associated equations of motion for the $56$ vector fields are twisted self-duality equations of motion without imposing further constraints, before adding the dynamics of the other fields of maximal supergravity to the system.

\begin{subsection}{Twisted self-duality in four space-time dimensions}\label{selfdual}
As a first step, we want to recall some basic facts about self-dual fields. In order for these to exist, two ingredients are necessary: Firstly, the field strength $\cF$ and its dual must have the same number of components and secondly, the square of the operation of taking the dual should give $+1$. The square of the usual Hodge dual $*$ on a two-form in four dimensional space-time, however, squares to $-1$, which rules out self-duality under the standard Hodge dual. Fortunately, maximal supergravity offers a different concept of duality that is intimately linked to its field content, which can be thought of as a twisted Hodge dual \cite{CJLP97}. Since the seventy scalars can be described by an $\cV\in E_{7(7)}/(SU(8)/\Z_2)$ coset \cite{CJ79}, the ``scalar metric'' $G=\cV\cV^T$ transforms as an $E_{7(7)}$-tensor.\footnote{The transposition in $G=\cV\cV^T$ is to be understood as acting on the matrix representation of the real Lie group $E_{7(7)}$ in terms of real $56 \times 56$ matrices as in \cite{CJ79}. Note in particular that the maximal compact subgroup of $E_{7(7)}$ (being isomorphic to $SU(8)/\Z_2$) is represented by orthogonal matrices.} As $E_{7(7)}$ furthermore is a subgroup of $Sp(56)$, the constant symplectic form $\Omega$ of the $56$-dimensional representation of $E_{7(7)}$ also is $E_{7(7)}$-covariant. The contraction of $G$ with the inverse symplectic form thus defines an almost complex structure $J$ acting on the $56$ dimensional fundamental representation of $E_{7(7)}$:\footnote{We use Einstein's summation conventions in this article and we denote the inverse of the symplectic form $\Omega_{\cMa\cNa}$ (of which we assume without loss of generality to have standard form $\Omega=\binom{\,\,0\,\,\,\,1}{-1\,\,0}$ as in \cite{H08}) by raising the indices, i.e. $\Omega^{\cMa\cPa}\Omega_{\cPa\cNa}=\delta^\cMa_\cNa$.}
\be\label{complex}
J^\cMa{}_\cNa&:=&\Omega^{\cMa\cPa}G_{\cPa\cNa}\\
\text{with}\quad J^\cMa{}_\cPa J^\cPa{}_\cNa&=&-\delta^\cMa_\cNa\nn\\
\text{and}\quad \cMa,\cNa,\cPa &=&1,\dots,56.\nn
\ee
Combining the Hodge dual with this scalar-dependent $J$-twist, we can write down a consistent self-duality equation for the field strength $\cF_{\mu\nu}{}^\cMa:=2\p_{[\mu}\cA_{\nu]}{}^\cMa$ of the $56$ vector fields $\cA_\mu{}^\cMa$:
\be\label{selfdualC}
\cF_{\mu\nu}{}^\cMa &=&-\frac{1}{2e_4}\e_{\mu\nu}{}^{\sigma\tau}J^\cMa{}_\cNa \cF_{\sigma\tau}{}^\cNa.
\ee
The space-time indices $\mu,\nu$ take values in $0,\dots,3$ and $e_4:=\det(-g)^{\frac12}$ is the standard volume element\footnote{Note that $\e^{\mu\nu\sigma\tau}$ is completely antisymmetric and it is normalized as $\e^{0123}=+1$. Its indices are lowered with the $4$-dimensional metric $g_{\mu\nu}$.} In the remaining part of this section, we shall construct the action functional $S_{\text{vec}}$ whose extremization with respect to the $56$ vector fields gives rise to the twisted self-dual equation (\ref{selfdualC}). In doing so, we can follow the construction of Henneaux and Teitelboim that is described in \cite{HT88}. Their first step consists of splitting both the field strength $\cF_{\mu\nu}{}^\cMa$ and the $d=4$ metric $g_{\mu\nu}$ into time and space in the standard way
\be\label{metric}
g_{\mu\nu}&=&\left(\begin{tabular}{cc} $-N^2 +\h_{ij}N^iN^j$ & $\h_{ij}N^j$\\ $\h_{ij}N^j$ & $\h_{ij}$\end{tabular}\right)_{\mu\nu}
\ee
with spatial indices $i,j=1,\dots,3$. The field strength $\cF_{\mu\nu}{}^\cMa$ is then decomposed into electric and twisted magnetic fields
\begin{subequations}\label{ELM}
\be
\cE_j{}^\cMa &:=& \cF_{0j}{}^\cMa -N^i\cF_{ij}{}^\cMa,\\
\cB_k{}^\cMa &:=& \frac{N}{2e_3}h_{kl}\e^{lij}J^\cMa{}_\cNa\cF_{ij}{}^\cNa,
\ee
\end{subequations}
where $\e^{ijk}$ is normalized as $\e^{123}=1$ and where $e_3:=\det(\h)^{\frac12}$ is the abbreviation for the spatial volume element, i.e. the square root of the determinant of the spatial metric $h_{ij}$ (\ref{metric}). The twisted self-duality equation (\ref{selfdualC}) is then equivalent to
\be\label{BWGL5d}
\cE_k{}^\cMa  &=& \cB_k{}^\cMa.
\ee
It is this equation of motion that we will obtain from our $E_{7(7)}$-invariant action $S_{\text{vec}}$. The action $S_{\text{vec}}$ is constructed from contractions of the electric with the twisted magnetic fields $\cE$ and $\cB$ in a non-standard way \cite{HT88}:
\be\label{action0}
S_{\text{vec}}[\cA_\mu{}^\cMa]&:=&\frac{1}{8}\int d^4x \frac{e_3}{N}\left(\cE_i{}^\cMa -\cB_i{}^\cMa\right)G_{\cMa\cNa}\h^{ij}\cB_j{}^\cNa.
\ee
This action functional $S_{\text{vec}}$ is manifestly invariant under gauge transformations
\be\label{gauge} 
\delta_g A_\mu{}^\cMa &=& \p_\mu \Lambda^\cMa.
\ee
The equations of motion are obtained by extremizing $S_{\text{vec}}$ with respect to the $56$ vector fields $\cA_\mu{}^\cMa$. One crucial observation at this stage is that the zero-component $\cA_0{}^\cMa$ drops out from the action $S_{\text{vec}}$, because its entire contribution to $S_{\text{vec}}$ is contained in a total derivative, which can be made manifest by substituting the definition of $\cE_j{}^\cMa$ (\ref{ELM}) into $S_{\text{vec}}$ and focussing on the $\cA_0{}^\cMa$ component for illustrational purpose (keeping in mind that $\Omega$ is a constant invariant tensor of $E_{7(7)}$):
\beg
\left.S_{\text{vec}}\right|_{\cA_0{}^\cMa}&=&\frac{1}{16}\int d^4x \p_k\cA_{0}{}^\cMa \Omega_{\cMa\cNa}
\e^{kij}\cF_{ij}{}^\cNa
\,=\,
\frac{1}{16}\int d^4x \p_k\left(\cA_{0}{}^\cMa \Omega_{\cMa\cNa}
\e^{kij}\cF_{ij}{}^\cNa\right)
.
\eeg
In other words, we will not alter the action $S_{\text{vec}}$ if we replace the electric field strength $\cE_j{}^\cMa$ in (\ref{action0}) by the $\cA_0{}^\cMa$-{\it independent} quantity 
\beg
\cEs_j{}^\cMa&:=&\cE_j{}^\cMa+\p_j\cA_0{}^\cMa\\
&\stackrel{(\ref{ELM})}{=}&\p_0\cA_{j}{}^\cMa -N^i\cF_{ij}{}^\cMa.
\eeg
Thus, the zero component $\cA_0{}^\cMa$ has disappeared completely from the action $S_{\text{vec}}$, which now reads
\be\label{action}
S_{\text{vec}}[\cA_i{}^\cMa]&=&\frac{1}{8}\int d^4x \frac{e_3}{N}\left(\cEs_i{}^\cMa -\cB_i{}^\cMa\right)G_{\cMa\cNa}\h^{ij}\cB_j{}^\cNa.
\ee
Note that the action functional $S_{\text{vec}}$ is still gauge invariant (\ref{gauge}), even though not in a manifest way. A short computation then leads to the following equation of motion for the remaining spatial components $\cA_i{}^\cMa$:
\be\label{BWGL5}
0\,\,=\,\,\frac{\delta S_{\text{vec}}}{\delta \cA_i{}^\cNa} 
&=&
\frac14\Omega_{\cNa\cMa}\e^{ijk}\p_j\left(\cEs_k{}^\cMa -\cB_k{}^\cMa\right).
\ee
Since the symplectic form $\Omega$ is constant, the equation of motion is equivalent to the statement that the differential one-form $\cEs-\cB$ is closed. Since we assume, as usual, trivial topology of the spatial slices of the $d=4$ manifold, every closed form is exact. Poincar\'e's lemma then implies that the equations of motion (\ref{BWGL5}) are equivalent to
\be\label{BWGL5b}
\cEs_k{}^\cMa -\cB_k{}^\cMa &=& \p_k v{}^\cMa,
\ee
where $v^\cMa$ is an arbitary function. Since the zero component $\cA_0{}^\cMa$ did not appear in the action $S_{\text{vec}}$ (\ref{action}), it is not a dynamical field of the theory a priori. Therefore, we can without loss of generality {\it define} $\cA_0{}^\cMa$ as being the function $v^\cMa$ entering (\ref{BWGL5b}), which transforms the equation of motion (\ref{BWGL5b}) into the expected form:
\be\label{BWGL5c}
\cE_k{}^\cNa  &=& \cB_k{}^\cNa.
\ee
Hence, we have succeeded in reproducing the twisted selfduality equation of motion (\ref{BWGL5d}), which contains the same information as the general covariant one stated in (\ref{selfdualC}). Before stating the complete bosonic action of maximal supergravity, we want to make some remarks:
\begin{itemize}
	\item The present analysis of a twisted self-duality equation on an arbitrary, four-dimensional Lorentzian manifold completely parallels the one of Henneaux and Teitelboim in \cite{HT88} in which they discussed self-dual $p$-forms in a Lorentzian manifold of dimension $d=2p+2$. Note that this result is a non-trivial extension of their procedure, because the scalar metric $G_{\cMa\cNa}(x)$ (which is not constant in contradistinction to the symplectic form $\Omega$) is an essential ingredient in defining the twisted self-duality.
	\item At a first glance, the counting of degrees of freedom appears not to match with supergravity, because a Kaluza--Klein reduction of $D=11$ supergravity leads to $28$ vector fields in $d=4$, each subject to {\textit{second-order}} equations and each having two on-shell degrees of freedom (like a photon). Thus, we arrive at $28\times 2$ degrees of freedom. In the present formulation, the $56$ vector fields obey {\textit{first-order}} equations, however, which require the same amount of initial data. Thus, the counting of the degrees of freedom matches. Yet another way to explain this agreement is the observation that the action $S_{\text{vec}}$ (\ref{action}) is based on the standard description for $56$ vector fields in $d=3$ Euclidean dimensions (that contain $56\times 1$ on-shell degrees of freedom), which are coupled to time in a particular way.
	\item The advantage of our approach is that the global $E_{7(7)}$-invariance is manifest for both the action and the equations of motion. Note that it is the guiding principle of this article to preserve the $E_{7(7)}$ symmetry. Only in the Appendix, where we will explicitly link the fields to $D=11$ supergravity, we will have to dispense with manifest $E_{7(7)}$-covariance for obvious reasons.
	\item It is interesting to observe that the occurrence of the potential $\cA_0{}^\cMa=v^\cMa$ in this procedure is completely analogous to the role played by the six-form potential $A_6$ in $D=11$ supergravity. In order to write the four-form equation of motion $d*F_4=\frac12F_4\wedge F_4$ in the first order form $*F_4=F_7$, it is necessary to introduce a six-form potential by $F_7=dA_6+\frac12A_3\wedge F_4$ \cite{CJLP98}.
\end{itemize}
\end{subsection}

\begin{subsection}{$E_{7(7)}$ invariant Lagrangian - bosonic part}
In the preceding section, we have used the space-time metric $g_{\mu\nu}$ and the scalar metric $G_{\cMa\cNa}$ for which we also have to specify the dynamics. Since we want to arrive at a theory with both general coordinate covariance and global $E_{7(7)}$ symmetry, it is natural to describe the dynamics of $g_{\mu\nu}$ by the Einstein--Hilbert action and the one of the scalars by the usual $\sigma$-model. Thus, we are led to the complete bosonic part of the action:\footnote{Note that the relative couplings are fixed in order to allow for a convenient comparison with $D=11$ supergravity, which is explained in detail in the Appendix.}
\be\label{action2}
S_{\text{bos}}&=& S_{\text{grav}} +S_{\text{scal}}+S_{\text{vec}}\nn\\
S_{\text{grav}} +S_{\text{scal}}
&=&\int e_4 d^4x \left[
\frac14 R
-\frac{1}{192}G^{\cMa\cNa}G^{\cPa\cQa}g^{\mu\nu}\p_\mu G_{\cMa\cPa}\p_\nu G_{\cNa\cQa}\right].
\ee
In order to obtain the equations of motion for the $70$ scalars $G_{\cMa\cNa}$, we first compute with $J^{\cMa}{}_\cPa J^{\cNa}{}_\cQa G_{\cMa\cNa}=G_{\cPa\cQa}$:
\be\label{BWGLG}
\frac{\delta S_{\text{vec}}}{\delta G_{\cMa\cNa}}\delta G_{\cMa\cNa}
&=&
-\frac{1}{16} e_4\h^{i_1j_1}\h^{i_2j_2}\cF_{i_1i_2}{}^\cMa\cF_{j_1j_2}{}^\cNa 
\delta G_{\cMa\cNa}.
\ee
Since the twisted self-duality equation of motion (\ref{selfdualC}) is not affected by the new terms, we can substitute it into this equation of motion (\ref{BWGLG}) to restore general coordinate covariance. In doing so, we use the fact $e_4=N e_3$ and that the scalars $\cV\in E_{7(7)}\subset Sp(56)$ form a symplectic matrix, which implies that the ``scalar metric'' $G=\cV\cV^T$ fulfills the relation $\delta G_{\cPa\cQa}=-J^{\cMa}{}_\cPa J^{\cNa}{}_\cQa \delta G_{\cMa\cNa}$:
\beg
\frac{\delta S_{\text{vec}}}{\delta G_{\cMa\cNa}}\delta G_{\cMa\cNa}
&\stackrel{(\ref{selfdualC})}{=}&
-\frac{1}{32} e_4
g^{\mu_1\nu_1}g^{\mu_2\nu_2}\cF_{\mu_1\mu_2}{}^\cMa\cF_{\nu_1\nu_2}{}^\cNa
\delta G_{\cMa\cNa}.
\eeg
Thus, the complete equation of motion of the scalars indeed shows general covariance:\footnote{Since the symmetric matrix $G_{\cMa\cNa}$ with $\cMa,\cNa=1,\dots,56$ is a highly redundant way to parametrize the $70$ scalars contained in $\cV\in E_{7(7)}/(SU(8)/\Z_2)$, we have kept the contraction with $\delta G_{\cMa\cNa}$ in eq. (\ref{BWGLG3}) which effectively enforces a projection on the terms in the parentheses. These projections will be made explicit in the following section.}
\be\label{BWGLG3}
0\,\,=\,\,\frac{\delta S_{\text{bos}}}{\delta G_{\cMa\cNa}}\delta G_{\cMa\cNa}
&=&
\Big[-\frac{1}{32} e_4
g^{\mu_1\nu_1}g^{\mu_2\nu_2}\cF_{\mu_1\mu_2}{}^\cMa\cF_{\nu_1\nu_2}{}^\cNa\nn\\
&&+\frac{1}{96}G^{\cMa\cPa}\p_\nu\left(e_4 g^{\mu\nu} G^{\cNa\cQa}\p_\mu G_{\cPa\cQa}\right)
\Big]\delta G_{\cMa\cNa}.
\ee
For the metric equation of motion, we face the complication that we have split the metric $g_{\mu\nu}$ into lapse $N$, shift $N^j$ and spatial metric $\h_{ij}$. This split (\ref{metric}) implies the identity
\be\label{Identg}
\frac{\delta S_{\text{vec}}}{\delta g_{\mu\nu}}
&=&
-\frac{1}{2N}\frac{\delta S_{\text{vec}}}{\delta N} \delta_0^\mu\delta_0^\nu
+\Big(\g^{ik}\frac{\delta S_{\text{vec}}}{\delta N^k}
+\frac{N^i}{N}\frac{\delta S_{\text{vec}}}{\delta N}
\Big)
\delta_0^{(\mu}\delta_i^{\nu)}\\
&&+\Big(
\frac{\delta S_{\text{vec}}}{\delta \g_{ij}}
-\g^{ik}\frac{\delta S_{\text{vec}}}{\delta N^k}N^j
-\frac{N^iN^j}{2N}\frac{\delta S_{\text{vec}}}{\delta N}
\Big)\delta_i^{(\mu}\delta_j^{\nu)}.
\nn
\ee
Substituting the twisted self-duality equation of motion (\ref{selfdualC}) in this expression then leads again to a covariant equation:
\beg
\frac{\delta S_{\text{vec}}}{\delta g_{\mu\nu}}
&=&
\frac{e_4}{16}G_{\cMa\cNa}g^{\rho\sigma}\cF_{\mu\rho}{}^\cMa \cF_{\nu\sigma}{}^\cNa.
\eeg

Thus, we arrive at the Einstein equation of motion
\be\label{BWGL4}
0\,\,=\,\,e_4^{-1}\frac{\delta S_{\text{bos}}}{\delta g_{\mu\nu}}
&=&
\frac14\left(\frac12g^{\mu\nu}R
-R^{\mu\nu}\right)
+
\frac{1}{16}G_{\cMa\cNa}g^{\rho\sigma}\cF_{\mu\rho}{}^\cMa \cF_{\nu\sigma}{}^\cNa\nn\\
&&+\frac{1}{192}G^{\cMa\cNa}G^{\cPa\cQa}\p^\mu G_{\cMa\cPa}\p^\nu G_{\cNa\cQa}\nn\\
&&-\frac{1}{384}g^{\mu\nu}G^{\cMa\cNa}G^{\cPa\cQa}\p_\sigma G_{\cMa\cPa}\p^\sigma G_{\cNa\cQa}.
\ee
Hence, all the equations of motion indeed show general covariance.
\end{subsection}

\begin{subsection}{Vielbein frame}\label{vielbeinS}
In order to couple the bosonic fields to fermions, we have to use a vielbein frame. The $d=4$ metric $g_{\mu\nu}$ (\ref{metric}) is written as
\be\label{vielbein}
g_{\mu\nu}&=& e_\mu{}^\alpha e_\nu{}^\beta\eta_{\alpha\beta}
\ee
with the Minkowski metric $\eta$ of signature $(-+++)$. Consistently with the decomposition of $g_{\mu\nu}$ into lapse, shift and spatial metric (\ref{metric}), we shall find it convenient to use a restricted frame of the form
 \be\label{gaugeviel}
 e_0{}^0&=& N\nn\\
 e_i{}^0&=& 0\nn\\
 e_0{}^a&=& e_i{}^a N^i\nn\\
 e_i{}^a&=& e_i{}^a.
 \ee
Furthermore, we want to rewrite the symmetric $E_{7(7)}$ tensor $G_{\cMa\cNa}$ in all expressions in terms of the coset $\cV\in E_{7(7)}/(SU(8)/\Z_2)$ using the following identification
\be\label{vielbeinG}
G_{\cMa\cNa}&=:&\cV_\cMa{}^{AB}\cV_{\cNa,AB} +\cV_{\cMa,AB}\cV_\cNa{}^{AB}\,\,=\,\,\cV_\cMa{}^{AB}\cV_{\cNa,AB} +\text{c.c.}
\ee
with $\cV_\cMa{}^{AB}=\cV_\cMa{}^{[AB]}$, where the indices $A,B=1,\dots,8$ label the $SU(8)/\Z_2$-representation of complex dimension $28$. As usual, complex conjugation changes the position of the $SU(8)$-indices, i.e. $(v^A)^*=v_A$. In complete analogy to the vielbein case, $\cV$ can be used to make the $E_{7(7)}$-index $\cMa$ ``flat'', e.g.
\be\label{flach1}
\cF_{\mu\nu}{}^{AB}&:=&\cV_\cMa{}^{AB}\cF_{\mu\nu}{}^\cMa.
\ee
This entails the following identity for the contraction of two arbitrary vectors $X^\cMa,Y^\cNa$ with the ``scalar metric'' $G$:
\be
G_{\cMa\cNa}X^\cMa Y^\cNa &=&X^{AB}Y_{AB} +\text{c.c.}
\ee
In analogy to the gravitational vielbein, we will distinguish the ``scalar vielbein'' $\cV\in E_{7(7)}/(SU(8)/\Z_2)$ from its inverse only by the different position of the ``curved'' indices $\cMa,\cNa,\ldots$, which implies together with $SU(8)$-covariance
\be
\cV_{AB}{}^\cMa\cV_\cMa{}^{CD}&=&\delta_{[A}^{[C}\delta_{B]}^{D]}\\
\cV_{AB}{}^\cMa\cV_{\cMa,CD}&=&0
\ee
and analogous statements for complex conjugated objects. Finally, the definition of the complex structure $J$ in (\ref{complex}) fixes the contraction of two arbitrary vectors $X^\cMa,Y^\cNa$ with the symplectic form $\Omega$ (up to a sign that we choose here):\footnote{Fur further details on the relation between $E_{7(7)}$-indices and $SU(8)$-indices, we refer the reader to \cite{H08}. An explicit example for this transformation can be found in eq. (\ref{ADec}) of the Appendix, which allows to verify the relation (\ref{symplectic}) for the symplectic form $\Omega$ which is of canonical form $\binom{\,\,0\,\,\,\,1}{-1\,\,0}$ \cite{H08}.}
\be\label{symplectic}
\Omega_{\cMa\cNa}X^\cMa Y^\cNa &=&iX^{AB}Y_{AB} +\text{c.c.}
\ee
Here, $i$ is the imaginary unit that satisfies $i^2=-1$. The fact that $\cV$ is a group element of $E_{7(7)}$ implies that its Maurer--Cartan form has the following property
\begin{subequations}\label{BA}
\be\label{BA1}
\cV_{AB}{}^\cMa\p_\mu \cV_\cMa{}^{CD}&=:&2(\cvB_{\mu})_{[A}{}^{[C}\delta_{B]}^{D]}\\
\cV_{AB}{}^\cMa\p_\mu \cV_{\cMa,CD}&=:&(\cvA_\mu)_{ABCD}\,\,=\,\,(\cvA_\mu)_{[ABCD]}\\
\text{with}\quad (\cvB_\mu)_A{}^A&=&0\nn.
\ee
\end{subequations}
Furthermore, the objects $\cvB$ and $\cvA$ are linked to their complex conjugates by
\begin{subequations}\label{BA2}
\be
(\cvA_\mu)_{ABCD}&=&\frac{1}{4!}\e_{ABCDEFGH}(\cvA_\mu)^{EFGH}\\
(\cvB_{\mu})_{A}{}^{C}&=& -(\cvB_{\mu})^{C}{}_{A}.
\ee
\end{subequations}
In other words, the $133$ dimensional Lie algebra $\mathfrak{e}_{7(7)}$ of $E_{7(7)}$ is split into the $63$ dimensional Lie algebra $\mathfrak{su}_8$ of $SU(8)/\Z_2$, parametrized by $\cvB_\mu$, and the $\binom{8}{4}=70$ dimensional representation of $\mathfrak{su}_8$. With this notation and the standard convention to use indices from the beginning of the alphabet $\alpha,A,\dots$ for the (flat) vielbein frame and indices from its middle $\mu,\cMa,\dots$ for the (curved) coordinate frame (e.g. $\p_\alpha=e_\alpha{}^\mu\p_\mu$), the three bosonic equations of motion (\ref{selfdualC}), (\ref{BWGLG3}) and (\ref{BWGL4}) take the following form:
\begin{subequations}\label{selfdualC2}
\be\label{selfdualC2a}
\cF_{\alpha_1\alpha_2}{}^{AB} \!\!&=&\!\!\!-\frac{i}{2}\e_{\alpha_1\alpha_2}{}^{\beta_1\beta_2}\cF_{\beta_1\beta_2}{}^{AB}\\
\frac{4}{3}\mathcal{D}_\alpha (\cvA^\alpha)_{ABCD}
    \!\!&=&\!\!\!
  \cF^{\alpha\beta}{}_{[AB}\cF_{\alpha\beta,CD]}+\frac{1}{4!}\e_{ABCDEFGH}\cF^{\alpha\beta,EF}\cF_{\alpha\beta}{}^{GH}
  \\
    R_{\alpha\beta} -\frac{1}{2}\eta_{\alpha\beta}R
   				\!\!&=&\!\!\!\frac{1}{4}
\big(\cF_{\alpha\g}{}^{AB}\cF_\beta{}^{\g}{}_{AB} 
 +\text{c.c.}\big)
 +\frac{1}{6}(\cvA_\alpha)_{ABCD}(\cvA_\beta)^{ABCD}
 \nn\\
   &&\!\!
  -\frac{1}{12}\eta_{\alpha\beta}(\cvA_\g)_{ABCD}(\cvA^\g)^{ABCD}.
  \label{selfdualC2c}
\ee
\end{subequations}
In the equation of the scalars, we have used the $SO(3,1)\times SU(8)/\Z_2$-covariant derivative $\mathcal{D}$ that is defined with the usual (Levi--Civita) spin connection $\omega$ and the $\mathfrak{su}_8$-valued connection $\cvB$ (\ref{BA}):
 \be\label{KovAbl}
 \mathcal{D}_\alpha (\cvA_\beta)_{ABCD}
 &=&
 \p_\alpha (\cvA_\beta)_{ABCD} +\omega_{\alpha\beta}{}^\g (\cvA_\g)_{ABCD}
 \nn\\
  &&-4(\cvB_\alpha)_{[A}{}^E (\cvA_\beta)_{BCD]E}.
 \ee
It is important to note that changing to the ``vielbein frame'' for the scalars does not violate the $E_{7(7)}$-covariance in the equations (\ref{selfdualC2}). We want to remark that this notation also suggests an easy comparison to the standard formulation of $d=4$ $\cN=8$ supergravity: To see this, we would have to violate the $E_{7(7)}$-covariance by splitting the complex $SU(8)$ representation $\cF_{\beta_1\beta_2}{}^{AB}$ into its real and imaginary part
\be
\cF_{\beta_1\beta_2}{}^{AB}&=&\Rea \left(\cF_{\beta_1\beta_2}{}^{AB}\right) +i\Ima \left(\cF_{\beta_1\beta_2}{}^{AB}\right)
\ee
that constitute only $SO(8)$ representations and taking these as independent objects. The self-duality equation of motion in the form (\ref{selfdualC2a}) would then be equivalent to
\be\label{ImRe}
\Ima \left(\cF_{\alpha_1\alpha_2}{}^{AB}\right) &=&-\frac{1}{2}\e_{\alpha_1\alpha_2}{}^{\beta_1\beta_2}\Rea \left(\cF_{\beta_1\beta_2}{}^{AB}\right).
\ee
This would allow us to substitute the imaginary part in the other two equations by the real part. The latter could be identified with the field strength of the $28$ vector fields that arise from a Kaluza--Klein reduction of $D=11$ supergravity, which form an $SO(8)$ representation \cite{CJ79,dWN82,CJLP97}. We have checked that the resulting equations of motion completely coincide with the ones of maximal supergravity \cite{CJ79,dWN82}. We will provide the explicit relations between the $D=11$ quantities and the field strengths $\cF_{\alpha_1\alpha_2}{}^{AB}$ as well as details of this check in the Appendix. At this point, we only want to remark that the twisted self-duality equation of motion provides both the Bianchi identity and the (generalized) Maxwell equation of motion for the remaining $28$ vector fields. \\

A last comment concerns the constants of normalization. The relative coupling in the action $S$ (\ref{action2}) has been chosen in such a way that the identification with $D=11$ supergravity is as simple as possible. It is important to note however that the matching of the equations of motion for both the metric and the scalars with supergravity is not due to a suitable choice of normalization, but indeed contains non-trivial information. This non-trivial coupling is in fact fixed by the Chern--Simons term of $D=11$ supergravity. A full explication of these facts can be found in the Appendix.
\end{subsection}

\end{section}

\begin{section}{Coupling to the fermions}\label{fermions}
\begin{subsection}{Fermionic action}
As a next step, we will couple the bosonic action $S$ (\ref{action2}) to fermions. As usual, these form representations of the covering of the Lorentz group, in this case of $\Spin(3,1)\times SU(8)$. The Weyl spinors of $d=4$ $\cN=8$ supergravity constitute the $56$ dimensional representation $\chi^{ABC}=\chi^{[ABC]}$ of $SU(8)$ and the gravitini the $8$ dimensional one denoted by $(\chi_\mu)^A$ with $A,B,C=1,\dots,8$. Since $D=11$ supergravity is stated in terms of Majorana spinors, we will also adopt this notation by using chiral Dirac spinors. In formul\ae, we will use the Majorana representation of the Clifford algebra in $d=4$ with the Minkowski metric $\eta$ of signature $(-+++)$:
\be\label{clifford}
\{\g^\alpha,\g^\beta\}&=&2\eta^{\alpha\beta}\\
\text{with}\quad \g_5\e^{\alpha\beta\g\delta}&:=&\g^{\alpha\beta\g\delta}\,\,=\,\,\g^{[\alpha}\g^{\beta}\g^{\g}\g^{\delta]}.
\nn
\ee
This implies $\g_5^2=-\id$ and that all $\g$-matrices are real. The chiral spinors $\chi^{ABC}$ and $(\chi_\mu)^A$ are then subject to the constraint $\g_5\chi^{ABC}=i\chi^{ABC}$ and $\g_5(\chi_\mu)^A=i(\chi_\mu)^A$ with the imaginary unit $i$ already used for the symplectic form $\Omega$ in (\ref{symplectic}).\footnote{In other words, acting with the projector $P^+=\frac12(\id+i\g_5)$ on the chiral spinor $\chi^{ABC}$ is trivial $P^+\chi^{ABC}=0$, which is the well-known formulation used in e.g. \cite{CJ79,dWN86}.} Complex conjugation of the $SU(8)$ representation amounts to lowering the $SU(8)$-indices and hence we obtain by consistency e.g. $\g_5\chi_{ABC}=-i\chi_{ABC}$.\footnote{The Majorana conjugation $\check{\chi}^{ABC}:=(i\g^0\chi^{ABC})^T$ with the transposition acting on the spinor indices does not affect the $SU(8)$-indices. This is why we refrain from using the notation $\bar{\chi}$ for conjugated spinors that is conventionally understood to also include a complex conjugation, which is not the case here. Furthermore, we use the standard convention for the complex conjugation of classical fermions $\chi_1,\chi_2$, i.e. that $i\chi_1^T\chi_2$ is real.} With these conventions, it is straightforward to obtain the fermionic part of the action by a Kaluza--Klein reduction of $D=11$ supergravity as we explain in detail in the Appendix. We will state the action at first and then explain the subtlety in the coupling to the $56$ vector fields:\footnote{We neglect the quartic fermionic contributions to the action at this stage. We will comment on their inclusion in section \ref{nonlinear}.}
\be\label{actionferm}
	S_{\text{ferm}}
  &=& \int e_4 d^4 x 
 \left\{
		-\frac{1}{2}(\check{\chi}_{\beta})^A\g^{\beta \gamma\delta}e_\delta{}^\mu
\mathcal{D}_\gamma(\chi_{\mu})_A
									-\frac{1}{96}\check{\chi}^{ABC}\g^\gamma\mathcal{D}_\gamma \chi_{ABC}
		\right.\\
		&&
				-\frac{1}{12}\check{\chi}^{ABC}\g^\alpha\g^\beta(\chi_\alpha)^D(\cvA_\beta)_{ABCD}
																		\left.		-\frac{1}{4}\cW{}^{\beta_1\beta_2}{}_{AB}(P^-)_{\beta_1\beta_2}^{ab}\mathcal{F}_{ab}{}^{AB}
						+\text{c.c.}\right\}\nn
	\ee
	with the indices $a,b=1,\dots,3$ and $\cF_{ab}{}^{AB}= e_a{}^\mu e_b{}^\nu \cF_{\mu\nu}{}^{AB}=e_a{}^i e_b{}^j \cF_{ij}{}^{AB}$ due to the gauge fixing (\ref{gaugeviel}). Inside the action, we have used the $\Spin(3,1)\times SU(8)$-covariant derivative $\mathcal{D}$ that follows from eq. (\ref{KovAbl})\footnote{Following \cite{FN76}, the connection $\omega$ does not act on the vector index of the gravitino $\chi_\mu$ in eq. (\ref{Der4Db}). The action $S_{\text{ferm}}$ (\ref{actionferm}) is nevertheless $\Diff(4)$-invariant due to the antisymmetry $[\g\delta]$ in the first term on the r.h.s. of (\ref{actionferm}).}
	\begin{subequations}\label{Der4D}
	\be
	\mathcal{D}_\gamma \chi_{ABC}
	&:=&
	\p_\gamma \chi_{ABC}  +\frac14 \omega_{\gamma\beta_1\beta_2}\g^{\beta_1\beta_2}\chi_{ABC}
		+3(\cvB_\g)_{[A}{}^D \chi_{BC]D}
	\\
	\mathcal{D}_\gamma (\chi_\mu)_A
	&:=&
	\p_\gamma(\chi_\mu)_A  +\frac14 \omega_{\gamma\beta_1\beta_2}\g^{\beta_1\beta_2}(\chi_\mu)_A 
			+(\cvB_\g)_A{}^B  (\chi_\mu)_B .
	\label{Der4Db}
	\ee
	  \end{subequations}
	Furthermore, we introduced the abbreviation $\cW$ for the bifermionic quantity 
	\be\label{WpDefi2}
	\cW_{\beta_1\beta_2}{}^{AB}
	&:=&
	4(\check{\chi}_{[\beta_1})^A(\chi_{\beta_2]})^B	
		-(\check{\chi}_{[\beta_2})_C\g_{\beta_1]}\chi^{ABC}\nn\\
	&&
		-\frac{1}{4!^2}\check{\chi}_{CDE}\g_{\beta_1 \beta_2}\chi_{FGH}	\e^{ABCDEFGH}
	\ee
as well as the projector $P^-$:
\be\label{projector}
	(P^\pm)_{ \beta_1\beta_2}^{\beta_3 \beta_4}&:=&\frac12\left(
\delta_{ \beta_1\beta_2}^{\beta_3 \beta_4}
\pm\frac{i}{2}\e_{ \beta_1\beta_2}{}^{\beta_3 \beta_4}
	 	  	  	  \right).
	 	  	  	  \ee
Before discussing the equations of motion for the combined system $S=S_{\text{bos}}+S_{\text{ferm}}$ (\ref{action},\,\ref{actionferm}), we want to explain why we could not use the conventional way to couple the $28$ field strengths of $d=4$ $\cN=8$ supergravity to the fermions \cite{CJ79,dWN82}. Our new formulation is a necessary consequence in order to establish manifest $E_{7(7)}$-invariance in the Lagrangian. It is well-known in maximal supergravity theories that the fermions do not transform under the global symmetry group $E_{n(n)}$ with $n=11-d$ in $4<d<11$ dimensions \cite{dW02}, but only with respect to the covering of its compact subgroup. Note however that any bifermionic expression can be transformed into an $E_{n(n)}$-tensor by a contraction with the scalar coset matrix $\cV\in E_{n(n)}/K(E_{n(n)})$. Hence, it is sufficient for our purpose of stating a manifestly $E_{7(7)}$ invariant Lagrangian of maximal supergravity in $d=4$ that its fermionic part shows $SU(8)$-covariance. However, the field strength of the $28$ vector fields in the usual formulation of $d=4$ $\cN=8$ is not a viable object on the level of the Lagrangian, because it does not even form an $SU(8)$ representation off-shell. This is a first reason why we had to use the formulation involving the projector $P^-$ inside the fermionic action $S_{\text{ferm}}$ (\ref{actionferm}).\newpage

Another argument in favour of our $E_{7(7)}$-invariant formulation of the Lagrangian is related to supersymmetry. Given the bosonic action $S_{\text{bos}}$ (\ref{action}) that does not depend on the zero component $\cA_0{}^\cMa$ of the $56$ vector fields, the requirement of superinvariance $\delta S=0$ of the complete action $S=S_{\text{bos}}+S_{\text{ferm}}$ can only hold if $\cA_0{}^\cMa$ does not appear in $S_{\text{ferm}}$ either. This serves as a second argument for the statement that the standard formulation of the fermionic Lagrangian is not admissible in our case. To sum up our arguments, we are forced to break the manifest $\Diff(4)$-covariance in the fermionic Lagrangian as well in order to guarantee $E_{7(7)}$-invariance of the action functional $S$. Nevertheless, the equations of motion will show general covariance in complete analogy to the bosonic ones, as we shall verify next.
\end{subsection}

\begin{subsection}{Fermionic equations of motion}
For the comparison of the fermionic dynamics with maximal supergravity, it is important to keep in mind that we have obtained the fermionic action $S_{\text{ferm}}$ (\ref{actionferm}) from a Kaluza--Klein reduction of $D=11$ supergravity, apart from the terms involving the field strengths. Therefore, it is sufficient to discuss the equations that are affected by this change. We will proceed order by order in fermions. As a first step, we observe that the bosonic equations of motion (\ref{selfdualC2}) will only be modified by terms that are quadratic in fermions. This implies in particular that the twisted self-duality equation for the field strength $\cF$ in (\ref{selfdualC2a}) still holds to leading order in fermions. In view of this fact, we can substitute it again inside the equations of motion of both the gravitino $(\chi_\mu)^A$ and the Dirac spinor $\chi^{ABC}$ to restore manifest general covariance to leading order in fermions. We will address the complete theory including all orders in fermions $\chi$ in section \ref{nonlinear}, but at this point, we content ourselves with focussing on the leading order terms. Using left derivation, we obtain the two equations
\begin{subequations}\label{Schi}
\be\label{SchiABC}
	0&=&-e_4^{-1}\frac{\delta S}{\delta \check{\chi}^{ABC}}\\
  &=& \frac{1}{48}\g^\g\mathcal{D}_\g\chi_{ABC}
 		+\frac{1}{12}\g^\alpha\g^\beta(\chi_\alpha)^D(\cvA_\beta)_{ABCD}
		+\frac{1}{4} \g^{\beta_1}(\chi^{\beta_2})_{[A}(P^+)_{\beta_1\beta_2}^{ab}\cF_{ab}{}_{BC]} 
						\nn\\
		&&
												-\frac{1}{(4!)^2 2} \g^{\beta_1 \beta_2}\chi^{FGH}	\e_{ABCDEFGH}(P^-)_{\beta_1\beta_2}^{ab}\cF_{ab}{}^{DE} 
				+\mathcal{O}(\chi^3)
				\nn\\
					\label{Schinu}
	0&=&e_4^{-1}e_{\nu\beta}\frac{\delta S}{\delta (\check{\chi}_\nu)^{A}}\\
   &=& 
 		- \g_\beta{}^{ \gamma\delta} e_\delta{}^\mu
\mathcal{D}_\gamma(\chi_{\mu})_A
											+\frac{1}{12} \g^\alpha\g_\beta\chi^{BCD}(\cvA_\alpha)_{ABCD}
		\nn\\
		&&
		+2(\chi^{\alpha})^B	
		(P^+)_{\alpha\beta}^{ab}\cF_{ab}{}_{AB} 
									+\frac{1}{4} \g^{\alpha}\chi_{ABC}
					(P^-)_{\alpha\beta}^{ab}\cF_{ab}{}^{BC} 
					+\mathcal{O}(\chi^3).
	\nn\ee
	\end{subequations}
	As a next step, we can insert the twisted self-duality equation of motion of $\cF$ (\ref{selfdualC2a}) in the form
		\be\label{PFonshell}
		(P^-)_{\beta_1\beta_2}^{ab}\cF_{ab}{}^{AB}&\stackrel{!}{=}&\frac12\cF_{\beta_1\beta_2}{}^{AB} +\mathcal{O}(\chi^2)
		\ee
		as well as its complex conjugate in the equations (\ref{Schi}). Hence, the manifest $\Diff(4)$-covariance of the fermionic equations of motion is restored, too.\\
		
		In complete analogy to the bosonic equations of motion, we obtain perfect agreement with the equations of motion of $d=4$ $\cN=8$ supergravity in their usual form. To do this, we have to break the $SU(8)$-covariance by substituting the imaginary part of $\cF_{\beta_1\beta_2}{}^{AB}$ by its real part as done in eq. (\ref{ImRe}) (which is tantamount to imposing the twisted self-duality equation of motion). Then, we can identify $28$ field strengths with the usual ones of supergravity that can be obtained from a Kaluza--Klein reduction of $D=11$ supergravity as explained in the Appendix. Hence, we have shown that all equations of motion exhibit manifest general covariance and that they agree with maximal supergravity upon breaking the $SU(8)$ symmetry to $SO(8)$ to leading order in fermions $\chi$. Before commenting on the next-to leading-order contributions $\mathcal{O}(\chi^2)$ in section \ref{nonlinear}, we want to link bosons to fermions by supersymmetry and check the supersymmetry algebra as well as the superinvariance of the complete action $S$ to leading order in fermions.
\end{subsection}

\begin{subsection}{Supersymmetry}	
The supersymmetry variations of the dynamical fields can in principle be derived from a Kaluza--Klein reduction of the ones of $D=11$ supergravity on a flat seven-torus (in complete analogy to the action $S$). With the identifications stated in the Appendix, the supersymmetry transformations of the bosonic fields read
\begin{subequations}\label{SFerm3}
\be
e_\alpha{}^\mu \delta e_\mu{}^\beta &=& \check{\eps}^C\g^\beta(\chi_\alpha)_C +\text{c.c.}\\
\cV^{AB,\cMa}\delta \cV_\cMa{}^{CD} &=&\check{\eps}^{[A}\chi^{BCD]} + \frac{1}{4!}\e^{ABCDEFGH}\check{\eps}_E\chi_{FGH}\\
e_\alpha{}^\mu \cV_\cMa{}^{AB}\delta \cA_\mu{}^{\cMa}
&=&
-4\check{\eps}^{[A}(\chi_\alpha)^{B]} 
-\frac{1}{2}\check{\eps}_{C}\g_\alpha\chi^{ABC}.
\label{SFerm3c}
\ee
\end{subequations}
In this analysis, there is a subtlety with the vector fields, as expected. Since only $28$ of these can be deduced from $D=11$ supergravity, we do not obtain the complete equation (\ref{SFerm3c}) from a simple Kaluza--Klein reduction. To be precise, we obtain the r.h.s, which is $SU(8)$ invariant, but on the l.h.s., the summation over the index $\cMa$ is restricted to $1,\dots,28$ which reflects the lack of $28$ vector fields.\footnote{This statement can be verified explicitly with the formul\ae{} provided in the Appendix.} Therefore, it is natural in our formulation to extend equation (\ref{SFerm3c}) to an $E_{7(7)}$- or $SU(8)$-covariant one by adding the missing $28$ vector fields to the l.h.s. (thus completing the $56$ dimensional $E_{7(7)}$-representation). This is the form of the supersymmetry variation that we shall use for proving both the superinvariance of the action and the closure of the supersymmetry algebra. Note that we have also defined the supersymmetry variation $\delta$ of the {\textit{on-shell}} field $\cA_0{}^\cMa$ in (\ref{SFerm3c}), although it does not appear in the action functional $S_{\text{vec}}$ (\ref{action}). We continue with defining the variations of the fermions, that read to leading order in $\chi$
	\begin{subequations}\label{c23}
\be\label{c2G}
\delta\chi^{ABC}
&=&-4(\cvA_\beta){}^{ABCD}\g^\beta\eps_D 
+3\g^{ab}\cF_{ab}{}^{[AB}
\eps^{C]}
+\mathcal{O}(\chi^2)\\
\label{c3G}
\delta(\chi_\mu)^A
&=& 
\mathcal{D}_\mu\eps^A
 +\frac{1}{4}\cF_{ab}{}^{AB}\g^{ab}\g_\mu\eps_{B}
 +\mathcal{O}(\chi^2)
\ee
\end{subequations}
with the $\Spin(3,1)\times SU(8)$-covariant derivative $\mathcal{D}$ already stated in (\ref{Der4D})
\be\label{supercovDG}
\mathcal{D}_\mu\eps^A
&:=&
\p_\mu \eps^A +\frac14\omega_{\mu\beta_1\beta_2}\g^{\beta_1\beta_2}\eps^A -(\cvB_\mu)_B{}^A\eps^B.
\ee
If we impose the equations of motion, the equations (\ref{c23}) exactly agree with the ones obtained from a Kaluza--Klein reduction of $D=11$ supergravity. Off-shell however, we have broken the manifest general covariance again in order to preserve $SU(8)$-covariance. [Note that the indices $a,b$ take the values $1,2,3$.] We have done this in complete analogy to the discussion of the fermionic action $S_{\text{ferm}}$ for the same reasons: since the zero component $\cA_0{}^{AB}$ does not appear in the bosonic action $S_{\text{bos}}$ (\ref{action}), it would be inconsistent to include it in either the fermionic Lagrangian or the supersymmetry variations of the fermions, because it would then prevent the complete action to be superinvariant, i.e. to fulfill $\delta S=0$. Nevertheless, the supersymmetry variations exhibit general covariance manifestly on-shell, which follows from imposing the twisted self-dual equation of motion of $\cF$ in the form (\ref{PFonshell}) and from the algebraic relations (\ref{Proj2}) stated in the Appendix.
\\

Before addressing the closure of the supersymmetry algebra, we have to come back to the supersymmetry variation on the seventy scalars that are parametrized by the coset $\cV\in E_{7(7)}/(SU(8)/\Z_2)$. In order for $\cV$ to describe only seventy off-shell degrees of freedom, we have to adopt some $SU(8)$-gauge fixing for the coset element $\cV$. It is now one of the basic structures of non-linear $\sigma$-models that a global left action (by $E_{7(7)}$) induces a local, compensating ($SU(8)/\Z_2$)-action that restores the gauge fixing of $\cV$. In general, this $SU(8)$-rotation depends on the fields of the coset $\cV$. Therefore, it is necessary to covariantize the supersymmetry variation. In other words, we have to modify the supersymmetry variation $\delta$ by a ``connection term'' (or a local $\mathfrak{su}_{8}$-transformation $\delta_\Sigma^{\mathfrak{su}_{8}}$) that exactly compensates the contribution that arises from a $\cV$-dependent $SU(8)$-rotation. (This necessity has already been observed in section 7.4 of \cite{CJ79}.) Starting from a general $\cV$ in any fixed gauge, the parameter $\Sigma$ of the $\mathfrak{su}_{8}$-transformation $\delta_\Sigma^{\mathfrak{su}_{8}}$ results from the projection of the $\mathfrak{e}_{7(7)}$ valued object $\cV^{-1}\delta \cV$ on $\mathfrak{su}_{8}$ in complete analogy to the Maurer--Cartan form (\ref{BA1}):
\be\label{SDelta1}
\cV_{CD}{}^\cNa \delta \cV_\cNa{}^{AB}&=:&2\Sigma^{[A}{}_{[C}\delta^{B]}_{D]}\\
 \Rightarrow\quad\Sigma^{A}{}_{C}&=&\frac13\cV_{CB}{}^\cNa \delta \cV_\cNa{}^{AB}.
 \nn
\ee
The covariant supersymmetry transformation on any $\mathfrak{su}_{8}$-representation is then defined by $\deltaS:=\delta  -\delta_\Sigma^{\mathfrak{su}_{8}}$ as in \cite{H08}. On the coset $\cV$, this leads to
\be\label{SDelta}
 \deltaS \cV_\cMa{}^{AB}&=&\delta \cV_\cMa{}^{AB} +2\Sigma^{[A}{}_{C} \cV_\cMa{}^{B]C}.
\ee
It is straightforward to verify that this covariantization $\deltaS$ of $\delta$ has no effect on neither the supersymmetry variations of the bosons (\ref{SFerm3}), nor on the ones of the fermions (\ref{c23}) to leading order in $\chi$, because $\Sigma$ is linear in fermions $\chi$ (\ref{SDelta1}).\footnote{Note that for the bosonic sector, it is possible to describe the coset in terms of the ``scalar metric'' $G_{\cMa\cNa}$ alone, on which both supersymmetry variations $\delta$ and $\deltaS$ yield the same result. For the coupling to the fermions however, we cannot dispense with the introduction of the ``vielbein frame'' $\cV$ and hence, we do have to include this additional $SU(8)$-action $\delta_\Sigma^{\mathfrak{su}_{8}}$ being the difference between $\delta$ and $\deltaS$.} For any gauge fixing of the coset $\cV \in E_{7(7)}/(SU(8)/\Z_2)$ however, we now obtain the further relation
\be\label{SDelta2}
\cV_{CD}{}^\cNa \deltaS \cV_\cNa{}^{AB}&=& 0.
\ee
Note that this procedure of ``covariantizing'' the supersymmetry transformation $\delta$ by combining it with a rotation is well-known \cite{dWN86} in Kaluza--Klein reductions, what we also illustrate in eq. (\ref{susy4}) of the Appendix.\footnote{We also want to mention that an explicit expression of $\Sigma$ in terms of $SU(8)$ covariant fermions must not exist, because $\Sigma$ is algebraic in fermions, but transforms as a connection. Nonetheless, the fact that $d=4$ $\cN=8$ supergravity in a block-triangular gauge is obtained from a Kaluza--Klein reduction of $D=11$ supergravity \cite{CJ79} allows to determine $\Sigma$ explicitly as a non-$SU(8)$-covariant algebraic function of the fermions of $D=11$ supergravity (eq. 4.5.22 in \cite{H08}).} \\

Equipped with these supersymmetry variations (\ref{SFerm3}), (\ref{c23}) and (\ref{SDelta2}), it is a brief computation to verify that the supersymmetry algebra closes on the bosons to leading order in $\chi$, if and only if we take into account the twisted self-duality equation of motion for $\cF$ (\ref{selfdualC2a}). Of particular interest is the commutator of two (covariant) supersymmetry variations which maps to a general coordinate transformation $\delta_{\Diff}$, a local Lorentz $\delta_{\mathfrak{so}_{(3,1)}}$ and a gauge transformation $\delta_{\text{gauge}}$ (where the latter only acts on the $56$ vector fields):
	\be\label{SUSYalg5}
\big[\deltaS_1,\deltaS_2\big]
	 	 	 	&=&
	 	 		\delta_{\Diff}
	 	 		 	 	+\delta_{\mathfrak{so}_{(3,1)}}
	 	 	 	+\delta_{\text{gauge}}
+\mathcal{O}(\chi^2).
	 	\ee
It is indeed this structure that results ``on-shell'' from an evaluation of the commutator of two supersymmetry variations on the bosonic fields $e_\mu{}^\alpha$, $\cV_\cMa{}^{AB}$ and $\cA_\mu{}^{\cMa}$:
	\begin{subequations}\label{SUSY}
	\be
\big[\deltaS_1,\deltaS_2\big]e_\mu{}^\alpha
	 	 	 	&=&
	 	 	\xi^\nu \p_\nu e_\mu{}^\alpha
	 	 		 	 	+e_\nu{}^\alpha \p_\mu \xi^\nu
	 	 		 	 	+\Sigma^\alpha{}_\beta e_\mu{}^\beta
	 	 		 	 	+\mathcal{O}(\chi^2)\\
	 	 		 	 	\big[\deltaS_1,\deltaS_2\big]\cV_\cMa{}^{AB}
	 	 	 	&=&
	 	 	\xi^\nu \p_\nu \cV_\cMa{}^{AB}
	 	 		 	 	+\mathcal{O}(\chi^2)\\
	 	 		 	 	\big[\deltaS_1,\deltaS_2\big]\cA_\mu{}^{\cMa}
	 	 	 	&\stackrel{(\ref{selfdualC2a})}{=}&
	 	 	\xi^\nu \p_\nu \cA_\mu{}^{\cMa}
	 	 		 	 	+\cA_\nu{}^{\cMa}\p_\mu \xi^\nu
	 	 		 	 	+\p_\mu\Lambda^\cMa
	 	 		 	 	+\mathcal{O}(\chi^2).
	 	 		 	 		 	\ee
 \end{subequations}
 For the closure to hold on the $56$ vector fields $\cA_\mu{}^\cMa$, it is essential to impose the twisted self-duality equation of motion for $\cF$ (\ref{selfdualC2a}). Note that in these equations (\ref{SUSY}), the two supersymmetry parameters $\eps_1$ and $\eps_2$ have been combined into $\Diff(4)\times E_{7(7)}$-representations:
	\begin{subequations}
\be
\xi^\nu&:=&e_\alpha{}^\nu\check{\eps}_2^{A}\g^\alpha\eps^1_A +\text{c.c.}\\
\xi^\cMa&:=&-4\cV_{AB}{}^\cMa\check{\eps}_2^{A}\eps_1^B +\text{c.c.}
\ee
 \end{subequations}
 The second parameter $\xi^\cMa$ can be transformed into the $SU(8)/\Z_2$-frame in analogy to the procedure used for the field strength $\cF$ in (\ref{flach1}), i.e. $\xi^{AB}=\cV_\cMa{}^{AB}\xi^\cMa$. These bifermionic parameters $\xi$ are the building blocks of the $\mathfrak{so}_{(3,1)}$-parameter $\Sigma$ and the gauge parameter $\Lambda$ that have the form
	\begin{subequations}
\be
\Sigma_{\alpha\beta}&=&+\frac12 (P^+)_{\alpha\beta}^{ab}\cF_{ab,AB}\xi^{AB} +\text{c.c.}\\
\Lambda^\cMa&=& \xi^\cMa -\cA_\nu{}^{\cMa}\xi^\nu.
\ee
 \end{subequations}

It is well-known that the commutator of two supersymmetry variations (\ref{SUSYalg5}) (inside the supersymmetry algebra) also produces a local $SU(8)$-rotation $\delta_{\mathfrak{su}_{8}}$ as well as a supersymmetry variation $\deltaS'$ on the r.h.s. of (\ref{SUSYalg5}). These transformations are, however, of next-to-leading order $\mathcal{O}(\chi^2)$ in the fermions, which is the reason why we have suppressed them at this point. The terms proportional to $\mathcal{O}(\chi^2)$ also are important for the verification of the closure of the supersymmetry algebra on the fermions: Since the terms of order $\mathcal{O}(\chi^2)$ inside the supersymmetry variations of $\chi$ mix with the leading order ones in this computation, we refrain from discussing this question here. Instead, we content ourselves with pointing out that the closure of the supersymmetry algebra on the fermions should not deviate from the standard computation in $d=4$ $\cN=8$ supergravity \cite{CJ79,dWN82} for two reasons: Firstly, the present formulation of supergravity completely agrees with the standard one in the fermionic sector on-shell and secondly, our off-shell modifications inside the supersymmetry variation of the fermions (\ref{c23}) were uniquely fixed by the requirement of manifest $SU(8)$-covariance.\\

An aspect that does not immediately follow from a comparison with the Kaluza--Klein reduction of $D=11$ supergravity is the question whether our modification of the bosonic action is compatible with the requirement of superinvariance of the action $\delta S=0$. Due to the superinvariance of $d=4$ $\cN=8$ supergravity \cite{CJ79,dWN82}, it is in fact sufficient to check the terms that we have modified, i.e. all the terms that contain the vector fields $\cA_\mu{}^\cMa$. Inside the variation $\delta S=\deltaS S$, there are contributions linear in $\cA$ and others that are quadratic in $\cA$. We have checked explicitly to leading order in fermions that both types of terms cancel, hence implying $\delta S=0$. To see this, we note that the terms inside $\delta S$, which are linear in $\cA$, arise in the contributions
\beg
\left.\deltaS S\right|_{\text{linear in }\cA}
&=&
\left.\left(\deltaS \check{\chi}^{ABC}\frac{\delta S_{\text{ferm}}}{\delta \check{\chi}^{ABC}}
+\deltaS (\check{\chi}_\nu)^{A}\frac{\delta S_{\text{ferm}}}{\delta (\check{\chi}_\nu)^{A}} +\text{c.c.}\right)
+\frac{\delta S_{\text{vec}}}{\delta \cA_i{}^{\cMa}}\deltaS \cA_i{}^{\cMa}\right|_{\text{linear in }\cA}
\eeg
It is a straightforward computation to arrive at
\beg
&&\deltaS \check{\chi}^{IJK}\frac{\delta S_{\text{ferm}}}{\delta \check{\chi}^{IJK}}
+\deltaS (\check{\chi}_\nu)^{G}\frac{\delta S_{\text{ferm}}}{\delta (\check{\chi}_\nu)^{G}}|_{\text{linear in }\cA} +\text{c.c.}\\
&=&
\frac{e_4}{2}(P^+)_{\beta_1\beta_2}^{ab}\cF_{ab}{}_{AB} 
				e^{\mu\beta_2}\cV_{\cMa}{}^{AB} \p^{\beta_1}\deltaS\mathcal{A}_{\mu}{}^{\cMa} +\text{c.c.}\nn
				\eeg
 that completely agrees with $-\frac{\delta S_{\text{vec}}}{\delta \cA_i{}^{\texttt{m}}}\deltaS \cA_i{}^{\texttt{m}}|_{\text{linear in }\cA}$ up to a total derivative term, which guarantees the superinvariance of the action $S$ to linear order in $\cA$. To quadratic order in $\cA$, we can without loss of generality focus on the terms
\beg
\left.\deltaS S\right|_{\text{quadratic in }\cA}
&=&
\left(\left.\deltaS \check{\chi}^{ABC}\right|_{\cA}\left.\frac{\delta S}{\delta \check{\chi}^{ABC}}\right|_{\cA}
+\left.\deltaS (\check{\chi}_\nu)^{A}\right|_{\cA}\left.\frac{\delta S}{\delta (\check{\chi}_\nu)^{A}}\right|_{\cA} +\text{c.c.}\right)\\
&&+\frac{\delta S_{\text{vec}}}{\delta e_\alpha{}^\mu} \deltaS e_\alpha{}^\mu
+\frac{\delta S_{\text{vec}}}{\delta \cV_\cMa{}^{AB}} \deltaS \cV_\cMa{}^{AB}.
\eeg
 After some computation, we obtain modulo next-to-leading order terms $\mathcal{O}(\chi^2)$ in fermions:
\beg
&&\left.\deltaS \check{\chi}^{ABC}\right|_{\cA}\left.\frac{\delta S}{\delta \check{\chi}^{ABC}}\right|_{\cA}
+\left.\deltaS (\check{\chi}_\nu)^{A}\right|_{\cA}\left.\frac{\delta S}{\delta (\check{\chi}_\nu)^{A}}\right|_{\cA} +\text{c.c.}\\
&=&e_4(P^+)_{\g_1\g_2}^{ab}\eta^{\g_2\beta_1}(P^-)_{\beta_1\beta_2}^{cd}\cF_{ab}{}_{AB}\cF_{cd}{}^{AB}
			e{}^{\mu\beta_2} \delta e_\mu{}^{\g_1}\\
																&&+\frac{e_4}{8}\left(\cF_{}{}^{ab,CD}
					\cF_{ab}{}^{AB}
					\cV_{AB}{}^{\cMa}\deltaS \cV_{\cMa,CD}+\text{c.c.}\right)
								+\mathcal{O}(\chi^2).
			\eeg
 The second line agrees with $-\frac{\delta S_{\text{vec}}}{\delta e_\alpha{}^\mu}|_{\cA} \deltaS e_\alpha{}^\mu$ and the third one with $-\frac{\delta S_{\text{vec}}}{\delta \cV_\cMa{}^{AB}}|_{\cA} \deltaS \cV_\cMa{}^{AB}+\text{c.c.}$, which implies that the terms in $\delta S$ that are quadratic in $\cA$ also vanish.\footnote{In order to check the agreement of the third line with $-\frac{\delta S_{\text{vec}}}{\delta \cV_\cMa{}^{AB}}|_{\cA} \deltaS \cV_\cMa{}^{AB} +\text{c.c.}$, it is easiest to use the identity (\ref{Identg}) in order not to break the $SO(3,1)$-covariance for the supersymmetry algebra. Otherwise, one is forced to introduce a compensating $SO(3,1)$-rotation to the supersymmetry generator $\deltaS$ in view of the gauge fixing of the vielbein (\ref{gaugeviel}).} Hence, we have succeeded to prove the superinvariance of the action $S$ to leading order in fermions. We will address the next-to-leading order terms in fermions in the following section.
\end{subsection}

\begin{subsection}{Non-linear contributions in fermions}\label{nonlinear}
Summarizing our results so far, we have shown that it is possible to construct an $E_{7(7)}$-invariant action $S=S_{\text{bos}} +S_{\text{ferm}}$ (\ref{action2},\,\ref{actionferm}) together with $E_{7(7)}$- or resp. $SU(8)$-covariant supersymmetry variations (\ref{SFerm3},\,\ref{c23}) for which the supersymmetry algebra (\ref{SUSYalg5}) closes and that leave the action $S$ invariant, i.e. $\delta S=\deltaS S=0$, to leading order in fermions $\chi$. Furthermore, our $E_{7(7)}$-covariant formulation of $d=4$ $\cN=8$ supergravity is completely equivalent to the standard approach, because both the supersymmetry variations and the equations of motion agree upon an (``on-shell'') elimination of $28$ vector fields (\ref{ImRe}). Therefore, it is natural to expect that the given theory coincides with maximal supergravity in $d=4$, including the next-to-leading order contributions in fermions that appear in both the action and the supersymmetry variations of the fermions of maximal supergravity \cite{CJ79,dWN82}.\\

In particular, given the closure of the supersymmetry algebra in $d=4$ $\cN=8$ supergravity and the superinvariance of its action \cite{CJ79,dWN82}, it is completely sufficient for the proof to focus again on the terms that we have modified in order to obtain manifest $E_{7(7)}$-covariance. In this context, an immediate question that arises concerns the possibility of a bifermionic coupling to the twisted self-duality equation of motion of the field strengths $\cF$ (\ref{selfdualC2a}). This will however not lead to any complications, on the contrary, it leads to the most natural generalization of this equation of motion, namely to a twisted self-duality of the {\it supercovariant} field strength $\hat{\cF}$ that we shall define in (\ref{Fsplit}) below.\\

To show this, we can largely follow the procedure of section \ref{selfdual} that led to the twisted self-duality equation of motion. The fact that the time component $\cA_0{}^\cMa$ of the vector fields appears in the action $S=S_{\text{bos}}+S_{\text{ferm}}$ (\ref{action2},\,\ref{actionferm}) only as a total derivative is not altered by the inclusion of the fermions. Hence, $\cA_0{}^\cMa$ does not provide an equation of motion. Therefore, it is sufficient to focus on the spatial components $\cA_i{}^\cMa$ as in eq. (\ref{BWGL5}).  It is a straightforward exercise to include the contribution from the fermionic action $S_{\text{ferm}}$ (\ref{actionferm}) in the variation with respect to $\cA_i{}^\cMa$, which generalizes eq. (\ref{BWGL5}) to
\be\label{BWGL6}
0\,\,=\,\,\frac{\delta S}{\delta \cA_i{}^\cMa} 
&=&
\frac14\Omega_{\cMa\cNa}\e^{ijk}\p_j\left(\cEs_k{}^\cNa -\cB_k{}^\cNa 
-T_k{}^\cNa
\right),
\ee
where the bifermionic quantity $T_k{}^\cNa$ is linked to the expression $\cW$ from (\ref{WpDefi2}) by
\beg
T_{k}{}^\cNa
&:=&
Ne_{kc}\e^{abc} \Omega^{\cNa\cMa}\cV_\cMa{}^{AB}(P^-)^{\beta_1\beta_2}_{ab}\cW{}_{\beta_1\beta_2}{}_{AB}
 +\text{c.c.}
\eeg
Using exactly the same analysis as in section \ref{selfdual}, we can transform the second order equation of motion (\ref{BWGL6}) into a first order one by identifying the resulting exact form (arising in the integration) with the time component of the $56$ vector fields. Thus, we arrive at the following generalization of (\ref{BWGL5c}):
\be\label{BWGL6c}
\cB_k{}^\cNa 
&=&
\cE_k{}^\cNa 
-T_k{}^\cNa.
\ee
The bifermionic expression $T$ on the r.h.s. does precisely have the correct shape to restore general covariance. After reexpressing the electric and the twisted magnetic field strengths $\cE$ and $\cB$ by the ordinary one $\cF$ (\ref{ELM}), we can state the generalization of the twisted self-duality equation of motion (\ref{selfdualC}), using the $SU(8)$-frame for convenience:
\be\label{selfdualC3}
\hat{\cF}_{\alpha_1\alpha_2}{}^{AB} &=&-\frac{i}{2}\e_{\alpha_1\alpha_2}{}^{\beta_1\beta_2}\hat{\cF}_{\beta_1\beta_2}{}^{AB}.
\ee
We are using the abbreviation
\be\label{Fsplit}
	\hat{\cF}_{\alpha_1\alpha_2}{}^{AB}&:=&\cF_{\alpha_1\alpha_2}{}^{AB} + \cW_{\alpha_1\alpha_2}{}^{AB}
	\ee
which can be checked with the supersymmetry variations of bosons and fermions (\ref{SFerm3},\,\ref{c23}) to be {\it supercovariant}. The latter fact could have been expected, because the superinvariance of the action guarantees that a supersymmetry variation of the bosonic equations of motion is proportional to the fermionic ones and vice versa.\\

We want to continue with a further statement concerning the next-to-leading order terms in fermions $\chi$ inside the supersymmetry algebra. We have checked explicitly that the supersymmetry algebra closes on the bosons to all orders in fermions, if and only if the first order equation $\hat{\cF}_{\alpha_1\alpha_2}{}^{AB} =-\frac{i}{2}\e_{\alpha_1\alpha_2}{}^{\beta_1\beta_2}\hat{\cF}_{\beta_1\beta_2}{}^{AB}$ (\ref{selfdualC3}) is imposed. As we have already hinted at above, the commutator of two supersymmetry variations $\deltaS_1$ and $\deltaS_2$ generates another supersymmetry variation $\deltaS'$ and an $SU(8)$-rotation $\delta_{\mathfrak{su}_{8}}$. This modifies the algebra (\ref{SUSYalg5}) to the standard form for supergravity theories 
\be\label{SUSYalg5b}
\big[\deltaS_1,\deltaS_2\big]
	 	 	 	&=&
	 	 		\deltaS'
	 	 		+\delta_{\Diff}
	 	 			 	 		 	 	+\delta_{\mathfrak{so}_{(3,1)}}
	 	 	+\delta_{\mathfrak{su}_{8}}
	 	 	+\delta_{\text{gauge}}
	 	\ee
which only closes ``on-shell'', i.e. modulo terms proportional to the equations of motion. The additional supersymmetry variation $\deltaS'$ on the r.h.s. has already been defined in \cite{dWN82}. In our conventions, it is determined by the parameter
\be\label{epsstrich}
\eps'{}^A&:=&-\xi^\nu(\chi_\nu)^A +\frac18\xi_{BC}\chi^{ABC}.
\ee
This directly extends all the three equations (\ref{SUSY}) to all orders in $\chi$ in such a way as to match the algebraic structure (\ref{SUSYalg5b}), given the supersymmetry variations of the fermions $\chi$ (\ref{c23}) are modified in the standard way \cite{dWN82}:
\begin{subequations}\label{c23G}
\be
\deltaS\chi^{ABC}
&=&-4(\hat{\cvA}_\beta){}^{ABCD}\g^\beta\eps_D 
+3\g^{ab}\hat{\cF}_{ab}{}^{[AB}
\eps^{C]}
\\
&&
-\frac{1}{48}\e^{ABCDEFGH}\eps^I
\big(\check{\chi}_{DEF}\chi_{GHI}\big)\nn\\
e_\alpha{}^\mu\deltaS(\chi_\mu)^A
&=& 
\hat{\mathcal{D}}_\alpha\eps^A
 +\frac{1}{4}\hat{\cF}_{ab}{}^{AB}\g^{ab}\g_\alpha\eps_{B}
 \\
  &&-\frac{1}{4!48}\e^{ABCDEFGH} \g^{\beta}\eps_H
 \big(\check{\chi}_{BCD}\g_{\alpha\beta}\chi_{EFG}\big)
 \nn\\
&&+\frac{1}{32} \g_\beta\g_\alpha\eps^D
\big(\check{\chi}^{ABC}\g^\beta\chi_{BCD}\big)
-\frac{1}{4} \g_\beta \eps_C
\big((\check{\chi}_\alpha)_B\g^\beta \chi^{ABC}\big),
\nn
\ee
\end{subequations}
where we have used the supercovariant quantity $\hat{\cvA}$
\be\label{superConf2}
(\hat{\cvA}_\beta){}^{ABCD}
&:=&
(\cvA_\beta){}^{ABCD}\\
&&-\Big((\check{\chi}_\beta)^{[A}\chi^{BCD]} +\frac{1}{4!}\e^{ABCDEFGH}(\check{\chi}_\beta)_{E}\chi_{FGH}\Big)\nn
\ee
and where the connection $\hat{\omega}$ in the supercovariant derivative $\hat{\mathcal{D}}$ differs from the one of $\mathcal{D}$ (\ref{supercovDG}) by a bifermionic contorsion contribution, namely
\be\label{superKont}
\hat{\omega}_{\alpha\beta_1\beta_2}
&:=&
\omega_{\alpha\beta_1\beta_2}
+
K_{\alpha\beta_1\beta_2}\nn\\
K_{\alpha\beta_1\beta_2}
&:=&
\frac12(\check{\chi}_{\beta_1})^C\g_\alpha(\chi_{\beta_2})_C
+(\check{\chi}_\alpha)^C\g_{[\beta_1}(\chi_{\beta_2]})_C
-\frac{1}{192}\check{\chi}^{ABC}\g_{\alpha\beta_1\beta_2}\chi_{ABC}
+\text{c.c.}
\nn
\ee
We want to emphasize again that the closure of the supersymmetry algebra of maximal supergravity on the fermions should not be affected by our modifications to the theory that only involved bosonic fields. We hence conclude that the consistency of the supersymmetry algebra of $d=4$ $\cN=8$ supergravity, as stated in \cite{dWN82}, implies the consistency of the supersymmetry algebra (\ref{SUSYalg5b}) in the fully $E_{7(7)}$-covariant form to all orders in fermions.\\

We close this section with a final remark on the proof of the superinvariance of the action $S$ to all orders in fermions. Given the superinvariance of the Kaluza--Klein reduction of $D=11$ supergravity and the agreement of the theories to leading order in fermions $\chi$, the non-linear terms in $\chi$ are uniquely fixed and can e.g. be taken from \cite{dWN82}. A complete review of $d=4$ $\cN=8$ supergravity is beyond the scope of this article, however.
		\end{subsection}
\end{section}

\begin{section}{Conserved Noether current}
After having constructed an $E_{7(7)}$-invariant Lagrangian that reproduces the equations of motion of maximal supergravity, it is a natural step to compute the conserved Noether current of the theory that is associated to this global symmetry. For the purely bosonic part of the theory defined by the action $S_{\text{bos}}$ (\ref{action2}), the standard procedure defines a conserved current for any constant $\Lambda$ in the matrix representation $\R^{56\times 56}$ of the Lie algebra $\mathfrak{e}_{7(7)}$ as follows:
\be\label{Noether1}
j_{\text{bos}}^\mu &:=&\left(2\frac{\delta S_{\text{bos}}}{\delta (\p_\mu G_{\cMa\cNa})} G_{\cMa\cPa} -\frac{\delta S_{\text{bos}}}{\delta (\p_\mu \cA_\nu{}^{\cPa})}\cA_\nu{}^\cNa
\right)\Lambda_\cNa{}^\cPa
\ee
with $\cMa,\cNa,\cPa=1,\dots,56$. Since the bosonic action $S_{\text{bos}}$ (\ref{action2}) does not depend on the zero-component $\cA_0{}^\cMa$, we could without loss of generality restrict in the definition of the current $j_{\text{bos}}^\mu$ (\ref{Noether1}) the summation over $\nu=0,\dots,3$ to $j=1,\dots,3$. This implies that all $\cA_0{}^\cMa$-dependences drop out of $j_{\text{bos}}^\mu$. For the discussion of symmetries however, the following form of the $\cA_0{}^\cMa$-{\textit{independent}} current will prove convenient:
\be\label{Noether2}
j_{\text{bos}}^\mu &=&\left(-\frac{e_4}{48} G^{\cNa\cMa}\p^\mu G_{\cMa\cPa} 
+\frac{1}{16}\e^{\mu\nu\rho\sigma}\cF_{\nu\rho}{}^\cMa\Omega_{\cPa\cMa}\cA_\sigma{}^\cNa\right.\\
&&\left.
+\frac18\delta^\mu_k\e^{ijk}\p_i\left(\Omega_{\cPa\cMa}\cA_j{}^\cMa\cA_0{}^\cNa\right)
-\frac{1}{4}\delta^\mu_k \e^{ijk} \cA_i{}^\cNa \Omega_{\cPa\cMa}\left(\cE_j{}^\cMa -\cB_j{}^\cMa\right)
\right)\Lambda_\cNa{}^\cPa.
\nn
\ee
The first line in (\ref{Noether2}) exhibits manifest general covariance, while the terms in the second line also have a special form: one is a curl and the other is proportional to the twisted self-dual equation of motion in the form $\cE_j{}^\cMa =\cB_j{}^\cMa$ (\ref{BWGL5c}). Therefore, it is an easy exercise to verify explicitly that the divergence of $j_{\text{bos}}^\mu$ vanishes on-shell for any $\Lambda\in \mathfrak{e}_{7(7)}$. In other words, the equations of motion of the $56$ vector fields $\cA_\mu{}^\cMa$ (\ref{selfdualC}) and the ones for the scalars $G_{\cMa\cNa}$ (\ref{BWGLG3}) guarantee that the $\mathfrak{e}_{7(7)}$-valued Noether current $j_{\text{bos}}^\mu$ (\ref{Noether2}) for the purely bosonic theory defined by the action $S_{\text{bos}}$ (\ref{action2}) is conserved:
\be\label{Noether3}
\p_\mu j_{\text{bos}}^\mu&=&0.
\ee
In order to couple the bosonic theory to fermions as performed in section \ref{fermions}, it was of crucial importance to switch to the vielbein frame for both the space-time metric $g_{\mu\nu}$ (\ref{vielbein}) and the ``scalar metric'' $G_{\cMa\cNa}$ (\ref{vielbeinG}). This implies that we have to substitute $G_{\cMa\cNa}$ for $\cV_\cMa{}^{AB}$ in the definition of the Noether current (\ref{Noether1}). Hence, for the complete theory defined by the action $S=S_{\text{bos}}+S_{\text{ferm}}$ (\ref{action2},\,\ref{actionferm}), the Noether current for any constant $\Lambda\in \mathfrak{e}_{7(7)}$ reads:
\be\label{Noether4}
j^\mu &:=&\left(\left(\frac{\delta S}{\delta (\p_\mu \cV_{\cNa}{}^{AB})} \cV_{\cPa}{}^{AB}+\text{c.c.}\right) -\frac{\delta S}{\delta (\p_\mu \cA_\nu{}^{\cPa})}\cA_\nu{}^\cNa
\right)\Lambda_\cNa{}^\cPa.
\ee
With the abbreviations (\ref{BA}) and (\ref{Der4D}), it is a straightforward computation to arrive at
\be\label{Noether7}
j^\mu &=&j_{\text{bos}}^\mu +j_{\text{ferm}}^\mu
\ee
with the bifermionic contribution
\beg
j_{\text{ferm}}^\mu
&=&
 e_4 
 \left\{
		\frac{1}{6}e_\g{}^\mu\left[
		(\check{\chi}_{\beta})^A\g^{\beta \gamma\delta}
(\chi_{\delta})_{E} \cV_{FA}{}^\cNa
									+\frac{1}{16}\check{\chi}^{ABC}\g^\gamma \chi_{BCE}\cV_{FA}{}^\cNa
		\right.		\right.\\
		&&\left.
				-\frac{1}{2}\check{\chi}_{[ABE}\g^\alpha\g^\g(\chi_\alpha)_{F]}\cV^{AB,\cNa}
				\right]\cV_\cPa{}^{EF}
								\nn\\
								&&
		\left.		+\frac{1}{2}\cW{}^{\beta_1\beta_2}{}_{AB}(P^-)_{\beta_1\beta_2}^{ab}e_a{}^{\mu}e_b{}^{j}\cV_\cPa{}^{AB}\cA_j{}^\cNa
						+\text{c.c.}\right\}\nn
\Lambda_\cNa{}^\cPa.
\eeg
The parametrization of the $\mathfrak{e}_{7(7)}$ valued constant $\Lambda$ by a $56\times 56$-matrix is highly redundant, in complete analogy with the description of the $70$ scalars by the symmetric matrix $G_{\cMa\cNa}$ in eq. (\ref{BWGLG3}). The separation of the constant $133$-dimensional parameter $\Lambda\in \mathfrak{e}_{7(7)}$ from the current $j^\mu$ hence also implies that the coefficients $(j^\mu)^\cNa{}_\cPa$ in $j^\mu =: (j^\mu)^\cNa{}_\cPa \Lambda_\cNa{}^\cPa$ are subject to various projection identities. The standard way to make these projections manifest would be to decompose $\Lambda_\cNa{}^\cPa$ into its linearly independent degrees of freedom in analogy to the treatment of the ``scalar metric'' $G_{\cMa\cNa}$ in section \ref{vielbeinS}. Since this does not provide any new insights at this point, we refrain from stating the explicit formul\ae{} here, but we will address this point in more detail in section \ref{CNoet}. Following the Noether theorem, the complete Noether current is conserved:
\beg
0&=&\p_\mu (j^\mu)^\cNa{}_\cPa.
\eeg
It is also interesting to observe that the quartic terms in fermions that we neglected in the fermionic part of the action $S_{\text{ferm}}$ do not provide any contribution to the Noether current, because these do neither contain the vectors $\cA_\nu{}^\cMa$ nor the scalars $G_{\cMa\cNa}$ \cite{dWN82}. Therefore, we may conclude that the conserved current $j^\mu$ (\ref{Noether7}) is exact to all orders in fermions $\chi$.\\

It has already been noticed by Gaillard and Zumino \cite{GZ81} that an $E_{7(7)}$-current $j^\mu$ cannot be invariant under gauge transformations $\delta_g \cA_\mu{}^\cMa=\p_\mu \Lambda^\cMa$, but that it has to transform in a slightly more general way, namely by the divergence of an antisymmetric tensor $M^{[\mu\nu]}$:
\be\label{trafo4}
j^\mu&\mapsto& j^\mu +\p_\nu M^{[\mu\nu]}.
\ee
This transformation does not alter the fact that the corresponding charge
\be\label{chargeE}
Q&:=&\int d^3x\,j^0
\ee
is gauge invariant, i.e. $Q\mapsto Q$. In our case, the property (\ref{trafo4}) of the current $j^\mu$ (\ref{Noether4}) is obvious, once the equations of motion are imposed. For the charge $Q$ however, a stronger statement follows from the associated current $j^\mu$ (\ref{Noether4}): $Q$ exhibits gauge invariance {\it independently} of the equations of motion. In order to show that $Q$ generates $\mathfrak{e}_{7(7)}$-transformations, we will have to pass to the Hamiltonian formalism.\\

Before performing this step in the next section, we acknowledge that a first result for the $E_{7(7)}$-Noether current $j^\mu$ was obtained in \cite{KS08} by using the Gaillard--Zumino approach, which consists of the following a posteriori procedure (without the complete knowledge of an $E_{7(7)}$-invariant action): Given the scalar and bifermionic contributions to the Noether current, the vector part of $j^\mu$ is defined in such a way that the resulting current is conserved. It is clear however that this approach can only reproduce an on-shell equivalent version of the Noether current $j^\mu_{\text{bos}}$ stated in eq. (\ref{Noether2}) up to an exact form, but not the complete object that is also defined off-shell in the present case. In particular, the Gaillard--Zumino approach is not satisfactory for the Hamiltonian analysis that we shall perform in the next section. 
\end{section}

\begin{section}{Hamiltonian formulation and general covariance}
In this section, we want to follow the analysis of Henneaux and Teitelboim to prove the general covariance of our system \cite{HT88}. To do this, we first switch to the Hamiltonian formalism and then we verify the hypersurface deformation algebra or Dirac algebra \cite{D64} between energy and momentum densities, which guarantees that the evolution from a given initial spacelike surface to a given final one is independent of the sequence of intermediate surfaces (employed to calculate the evolution). Furthermore, the gauge transformations corresponding to the (secondary) first-class constraints of vanishing energy and momentum densities will {\it define} the action of a diffeomorphism on the vector fields. Equipped with these transformations, we will finally verify that the action $S_{\text{bos}}$ (\ref{action2}) is invariant under a general coordinate transformation. Since the vector part $S_{\text{vec}}$ (\ref{action}) of the bosonic action is the one that does not show manifest $\Diff(4)$-covariance, we will focus on this first, before addressing the complete bosonic system.\footnote{We do not expect any complications to arise from the inclusion of the fermions in this analysis.}

\begin{subsection}{Vector part of the Hamiltonian}\label{vecHam}
In order to obtain the Hamiltonian associated to the vector part $S_{\text{vec}}$ (\ref{action}) of the bosonic action, we start from its Lagrangian density $\cL_{\text{vec}}$
\be\label{Lvec}
\cL_{\text{vec}}&:=&\frac{1}{8} \frac{e_3}{N}\left(\cEs_i{}^\cMa -\cB_i{}^\cMa\right)G_{\cMa\cNa}\h^{ij}\cB_j{}^\cNa.
\ee
It is straightforward to compute the conjugate momenta to the dynamical variables $\cA_i{}^\cMa$, where we will be using the standard abbreviation $\dot{f}$ for a time derivative $\p_0f$:
\be\label{pDefi}
\cpi^i_\cMa&:=&\frac{\p \cL_{\text{vec}}}{\p \dot{\cA}_i{}^\cMa}\nn\\
&=&
-\frac{1}{16} \Omega_{\cMa\cNa}
\e^{ijk}\cF_{jk}{}^\cNa
\ee
The momenta $\cpi^i_\cMa$ cannot be expressed as functions of the velocities $\dot{\cA}_i{}^\cMa$ on any spacelike surface due to the vanishing of the second derivative 
\beg
\frac{\p^2 \cL_{\text{vec}}}{\p \dot{\cA}_i{}^\cMa\p \dot{\cA}_j{}^\cNa}&=&0.
\eeg 
Following the terminology of Dirac \cite{D64}, we are left with the primary constraints relating the variables $\cA_i{}^\cMa$ to their conjugate momenta $\cpi^i_\cMa$ on any spacelike surface:
\be\label{constr}
\cPhi^i_\cMa&:=&\cpi^i_\cMa
+\frac{1}{8} \Omega_{\cMa\cNa}
\e^{ijk}\p_j\cA_{k}{}^\cNa.
\ee
The existence of these constraints implies that the Legendre transformation is singular in the following sense: A Hamiltonian $H$ uniquely determines a Lagrangian but not vice versa, because all Hamiltonians that differ by a linear combination of the constraints $\cPhi^i_\cMa(x)$ lead to the same Lagrangian. In other words, the Hamiltonian is only uniquely defined on the constraint hypersurface $\cPhi^i_\cMa=0$, for which we shall use the abbreviation $\cPhi=0$. There is however a way to solve this ambiguity and to single out a preferred Hamiltonian, which Dirac called the ``total Hamiltonian'' $H^{\text{vec}}_{\text{tot}}$, defined as follows in our case:
\be\label{Htotal}
H^{\text{vec}}_{\text{tot}}(\cA,\cpi) &:=&H^{\text{vec}}(\cA)|_{\cPhi=0} +\int d^3 x\, U(\cA,\cpi)_i^\cMa(x) \cPhi^i_\cMa(x).
\ee
The function $U(\cA,\cpi)_i^\cMa$ will be determined by requiring the primary constraints $\cPhi^i_\cMa$ (\ref{constr}) to be preserved under the time evolution (up to terms corresponding to gauge transformations). The Hamiltonian $H^{\text{vec}}(\cA)|_{\cPhi=0}$ is constructed from the usual Legendre tranformation of the Lagrangian $\cL_{\text{vec}}$ (\ref{Lvec}), in which we also impose the constraint $\cPhi=0$. This provides us with a function that turns out not to depend on the momenta $\cpi$ any more:\footnote{Note that for a more general system, the first term in (\ref{Htotal}) may also depend on the momenta \cite{D64}.}
\beg
H^{\text{vec}}(\cA)|_{\cPhi=0}&:=&\int d^3 x\, \cpi^i_\cMa(x) \p_0\cA_i{}^\cMa(x) -\cL_{\text{vec}}(x)\\
&=&\int d^3 x\, N(x) \cH(x)|_{\cPhi=0} + N^k(x) \cH_k(x)|_{\cPhi=0}.
\eeg
Here, the two densities $\cH^{\text{vec}}$ and $\cH^{\text{vec}}_k$ are defined on the constraint hypersurface $\cPhi=0$ by
\begin{subequations}\label{dense1}
\be
\cH^{\text{vec}}|_{\cPhi=0}&:=&\frac{e_3}{16}h^{i_1j_1}h^{i_2j_2}G_{\cMa\cNa}\cF_{i_1i_2}{}^\cMa\cF_{j_1j_2}{}^\cNa
\\
\cH^{\text{vec}}_k|_{\cPhi=0}&:=&-\frac{1}{16}\cF_{ki}{}^\cMa\cF_{j_1j_2}{}^\cNa\e^{ij_1j_2}\Omega_{\cMa\cNa}.
\ee
\end{subequations}
In order to fix the function $U(\cA,\cpi)$ in $H^{\text{vec}}_{\text{tot}}$ (\ref{Htotal}), we first define the equal time Poisson bracket for arbitrary functions $f,g$ of the variables $\cA,\cpi$ in the standard way:
\be\label{Poisson}
\{f,g\}_p&:=&\int d^3 x \left(\frac{\p f}{\p \cA_i{}^\cMa(x)}\frac{\p g}{\p \pi^i_\cMa(x)} -\frac{\p g}{\p \cA_i{}^\cMa(x)}\frac{\p f}{\p \pi^i_\cMa(x)}\right).
\ee
The Euler--Lagrange equations can then be stated as follows for any function $g$ of $\cA$ and $\cpi$ on the constraint hypersurface $\cPhi=0$ \cite{D64}:
\be
\dot{g} &=&\{g,H^{\text{vec}}_{\text{tot}}\}_p.
\ee
In particular, this equation determines the function $U$ by requiring
\be\label{bedingung}
0&=&\dot{\cPhi}^i_\cMa\,\,=\,\,\{\cPhi^i_\cMa,H^{\text{vec}}_{\text{tot}}\}_p
\ee
which can be evaluated using the Poisson bracket (\ref{Poisson}).\footnote{Note that we require this Poisson bracket (\ref{Poisson}) to vanish {\it independently} of the constraint hypersurface $\cPhi=0$. Otherwise the function $U$ would only be determined up to a linear combination of the constraints $\cPhi^i_\cMa$.} A special solution to this equation is
\be\label{special}
(U^s){}_i^\cMa&=& N^k\left(\cF_{ki}{}^\cMa 
-2\e_{ijk}\Omega^{\cMa\cNa}\cPhi_\cNa^j\right)\\
&&+\frac{N}{2\det(h)^{\frac12}}h_{ij}\left(J^\cMa{}_\cNa \e^{ji_1i_2}\cF_{i_1i_2}{}^\cNa +4 G^{\cMa\cNa}\cPhi_\cNa^j\right)
\nn.
\ee
The general solution of the equation (\ref{bedingung}) then is the sum of the special solution $U^s$ and the general solution $V$ of the associated homogeneous equation
\be\label{homog}
0&=&\int d^3x\, V_j^\cNa(x) \{\cPhi^i_\cMa(y),\cPhi^j_\cNa(x)\}_p\\
&=&\frac14\Omega_{\cMa\cNa}\e^{ijk}\p_jV_k{}^\cNa,
\nn
\ee
whose general solution is $V_k{}^\cNa=\p_k v^\cNa$ with arbitrary functions $v^\cNa(x)$ \cite{D64}. Thus, we have succeeded to construct the total Hamiltonian $H^{\text{vec}}_{\text{tot}}$ (\ref{Htotal}), which takes the form
\be\label{H1tot}
H^{\text{vec}}_{\text{tot}}
&=&\int d^3 x\, N \cH^{\text{vec}} + N^k \cH^{\text{vec}}_k +  v^\cMa \p_i\cPhi^i_\cMa.
\ee
The extension of the two densities $\cH^{\text{vec}}$ and $\cH^{\text{vec}}_k$ (\ref{dense1}) beyond the constraint hypersurface $\cPhi=0$ is hence uniquely fixed by eq. (\ref{bedingung}) to
\begin{subequations}\label{dense2}
\be
\cH^{\text{vec}}&=&\frac{1}{e_3}h_{ij}G^{\cMa\cNa}\left(
\cpi^i_\cMa \Omega_{\cPa\cNa}\e^{ji_1i_2}\p_{i_1}\cA_{i_2}{}^\cPa
+
2\cPhi_\cMa^i \cPhi_\cNa^j\right)
\\
\cH^{\text{vec}}_k&=&2\cpi^i_\cMa \p_{[k}\cA_{i]}{}^\cMa -2\e_{kij}\Omega^{\cMa\cNa}\cPhi_\cMa^i \cPhi_\cNa^j.
\ee
\end{subequations}

To describe this analysis in different words, the equation (\ref{bedingung}) guarantees that the primary constraints $\cPhi^i_\cMa$ (\ref{constr}) do not lead to secondary constraints. As a next step, we will explain why the functions $v^\cMa$ in the total Hamiltonian $H^{\text{vec}}_{\text{tot}}$ (\ref{H1tot}) are related to gauge transformations. This follows from the observation that not all the (primary) constraints $\cPhi^i_\cMa$ (\ref{constr}) are second-class, using Dirac's terminology from \cite{D64}. The functions $\p_i\cPhi^i_\cMa$ form a subset of first-class constraints. Hence, these generate gauge transformations $\delta_g$ of any function of phase space by taking the Poisson bracket. For the vector fields $\cA_i{}^\cMa$ this implies in particular with the definition of the constraints $\cPhi^i_\cMa$ (\ref{constr}) and the Poisson bracket (\ref{Poisson}):
\be\label{gaugeTrafo}
\delta_g \cA_i{}^\cMa (x)&=&\int d^3y\,\Lambda^\cNa(y) \{\p_j\cPhi^j_\cNa(y),\cA_i{}^\cMa(x)\}_p
\nn\\
&=&\p_i\Lambda^\cMa (x).
\ee
This is the transformation that we have introduced in eq. (\ref{gauge}) as an invariance of the action $S$. Furthermore, note that the link between first-class constraints and gauge tranformations (\ref{gaugeTrafo}) together with the relation $\{\cPhi^i_\cMa,H^{\text{vec}}_{\text{tot}}\}_p=0$ (\ref{bedingung}) also proves the statement that the Hamiltonian is invariant under gauge transformations, a fact already mentioned in \cite{GZ81}.\\

We want to phrase the gauge arbitrariness that we encountered in the definition of the total Hamiltonian in a different way in order to make contact to the Lagrangian analysis of section \ref{selfdual}. Given any initial data $(\cA,\cpi)$ on a spatial hypersurface $t=t_0$, the values of $(\cA,\cpi)(t)$ for $t>t_0$ are only determined up to gauge transformations, which correspond to different choices of the functions $v^\cMa(x)$ in the Hamiltonian $H^{\text{vec}}_{\text{tot}}$ (\ref{H1tot}) for $t=t_0$. In view of the $\Diff(4)$-covariant formulation of the equations of motion, it is natural to identify the functions $v^\cMa(x)$ in the Hamiltonian with the zero-component $\cA_0{}^\cMa$, which did not appear in the action $S$ nor the Hamiltonian a priori. Note that this procedure is in complete analogy to the analysis of the equations of motion that follow from the Lagrangian in (\ref{BWGL5b}). Furthermore, this choice is natural, because the first-class constraint $\p_j\cPhi^j_\cNa$ has the same form as the one for classical electrodynamics \cite{D64}. This analogy to electrodynamics will also prove to be useful for the verification of the invariance of our theory under general coordinate transformations. For this, we will also need some further ingredients from the standard Hamiltonian formulation of general relativity and of symmetric spaces that we shall summarize in the next section.
\end{subsection}

\begin{subsection}{Coupling to the scalars and to gravity}\label{coup}
We will prove the general covariance of our system by checking the hypersurface deformation algebra or Dirac algebra \cite{D64}. However, we have to keep in mind that it is inconsistent for the check of the algebra to restrict to the Hamiltonian densities $H^{\text{vec}}$ and $H^{\text{vec}}_k$ (\ref{dense2}), because these were computed from the vector part $S_{\text{vec}}$ (\ref{action}) of the bosonic action alone. Since both the scalars $G_{\cMa\cNa}$ as well as the space-time metric $g_{\mu\nu}$ appear in these densities (\ref{dense2}) explicitly, we have to include their corresponding Hamiltonians in the analysis.\footnote{The fermions, on the contrary, can be consistently dropped in the analysis of general covariance.}\\

Hence, we have to combine the vector Lagrangian $\cL_{\text{vec}}$ (\ref{Lvec}) with the terms describing the proper dynamics of the metric and the scalars, which are fixed by the action (\ref{action2}), or equivalently by its Lagrangian density:
\be\label{LaLL}
\cL_{\text{bos}}&=&\cL_{\text{vec}}
+ 
\frac{e_4}{4} R
-\frac{e_4}{192}G^{\cMa\cNa}G^{\cPa\cQa}g^{\mu\nu}\p_\mu G_{\cMa\cPa}\p_\nu G_{\cNa\cQa}.
\ee
We shall start with the Hamiltonian description of the seventy scalars that parametrize the symmetric space $E_{7(7)}/(SU(8)/\Z_2)$. For this, we have to keep in mind that their parametrization in terms of the symmetric tensor $G_{\cMa\cNa}$ with $\cMa,\cNa=1,\dots,56$ is highly redundant. In order to identify the independent phase space variables for the scalar sector, we will proceed in two steps:
\begin{enumerate}
	\item Starting from the definition $G=\cV\cV^T$ with $\cV\in E_{7(7)}/(SU(8)/\Z_2)$, we shall consider the $E_{7(7)}$-valued matrix $\cV_\cMa{}^{AB}$ (\ref{vielbeinG}) as an independent variable (with the $SU(8)$ indices $A,B=1,\dots,8$). This will define a $2\times 133$ dimensional sector of phase space.
	\item Since the Lagrangian only describes seventy physical scalars, it will give rise to the $SU(8)$-constraint, which imposes $63$ constraints on the momenta. Enforcing this weak equality then projects on the physical constraint hypersurface in phase space.
\end{enumerate}
In addressing the first point, we do not need to know the explicit form of the Lagrangian. It is sufficient to know that $\cL$ depends on the scalars only through an $E_{7(7)}$-valued matrix $\cV$ and its first derivatives. Focussing on the scalars and taking $\cV$ and its time derivative $\dot{\cV}$ as independent variables in the Lagrangian, we obtain modulo terms proportional to $\delta\cV$ and other fields:
\beg
\delta \cL|_\cV&=& \cpiV_{AB}^{\cMa}\delta \dot{\cV}_\cMa{}^{AB} + \text{c.c.} + \cO(\delta\cV)
\\
&=& \cpiV_{AB}^{\cMa}\cV_\cMa{}^{CD}\delta \left(\cV_{CD}{}^\cNa\dot{\cV}_\cNa{}^{AB}\right)
+\cpiV_{AB}^{\cMa}\cV_{\cMa,CD}\,\delta \left(\cV^{CD,\cNa}\dot{\cV}_\cNa{}^{AB}\right)
 + \text{c.c.}
 + \cO(\delta\cV).
\eeg
The fact that $\cV$ is $E_{7(7)}$-valued implies that the Maurer--Cartan form $\cV^{-1}\dot{\cV}$ takes values in the $133$ dimensional Lie algebra $\mathfrak{e}_{7(7)}$. Hence, the contractions of the momenta $\cpiV$ with $\cV$ must be constrained, too. We can make these restrictions explicit by defining ``new momenta'': 
\begin{subequations}\label{momenta}
\be
\cpiP_{ABCD}&:=&\cpiV_{[AB}^{\cMa}\cV_{\cMa,CD]} + \frac{1}{4!}\e_{ABCDEFGH}\cpiV^{EF,\cMa}\cV_\cMa{}^{GH}
\\
\cSU_A{}^B&:=&2\left(\delta_A^E\delta_C^B -\frac18\delta_A^B\delta_C^E\right)\left(\cpiV_{ED}^{\cMa}\cV_\cMa{}^{CD} - \cpiV^{CD,\cMa}\cV_{\cMa,ED}\right).
\ee
\end{subequations}
These are completely sufficient to write the variation of the Lagrangian $\delta \cL|_\cV$, which can be checked with the equations (\ref{BA}) and (\ref{BA2}):
\be\label{Lvar}
\delta \cL|_\cV
\,=\, \frac13\cSU_{[A}{}^{[C}\delta_{B]}^{D]}\delta \left(\cV_{CD}{}^\cNa\dot{\cV}_\cNa{}^{AB}\right)
+\cpiP_{ABCD}\,\delta \left(\cV^{CD,\cNa}\dot{\cV}_\cNa{}^{AB}\right)
 + \cO(\delta\cV).
\ee
To phrase this in other words, the original momentum $\cpiV_{AB}^{\cMa}$ is not a good phase space variable in general. Only for the case that the restrictions on $\cpiV_{AB}^{\cMa}$, i.e. the projections of $\cpiV\cV$ onto the Lie algebra $\mathfrak{e}_{7(7)}$, are imposed explicitly, one may use the naive Poisson brackets between $\cV$ and $\cpiV$, the only non-vanishing one being 
\beg
\{\cV_\cNa{}^{CD}(x),\cpiV_{AB}^{\cMa}(y)\}_p&=&\delta_\cNa^\cMa \delta_{[A}^{[C}\delta_{B]}^{D]} \delta^{(3)}(x-y)
.
\eeg
By construction, this condition is fulfilled for the new ``momenta'' $\cpiP_{ABCD}$ and $\cSU_A{}^B$. Hence, we can deduce their Poisson brackets from the ones of $\cV$ and $\cpiV$ directly, which leads to 
\be\label{Poisson3}
\left\{\cV_\cMa{}^{AB}(x),\cpiP_{EFGH}(y)\right\}_p&=&\cV_{\cMa,[EF}(x)\delta_{G}^A\delta_{H]}^B\delta^{(3)}(x-y)
\nn\\
\left\{\cV_{\cMa,AB}(x),\cpiP_{EFGH}(y)\right\}_p&=&\frac{1}{4!}\cV_{\cMa}{}^{CD}(x)\e_{ABCDEFGH}\delta^{(3)}(x-y)
\nn\\
\left\{\cV_\cMa{}^{AB}(x),\cSU_{E}{}^{F}(y)\right\}_p&=&-2\left(\delta_E^{[A}\cV_{\cMa}{}^{B]F}(x)+\frac18\cV_\cMa{}^{AB}(x)\delta_E^F\right)\delta^{(3)}(x-y)
\nn\\
\left\{\cV_{\cMa,AB}(x),\cSU_{E}{}^{F}(y)\right\}_p&=&2\left(\delta^F_{[A}\cV_{\cMa,B]E}(x)+\frac18\cV_{\cMa,AB}(x)\delta_E^F\right)\delta^{(3)}(x-y)
\nn\\
\left\{\cpiP_{ABCD}(x),\cpiP^{EFGH}(y)\right\}_p&=&\frac{2}{3}\delta_{[A}^{[E}\delta_B^F\delta_C^G\cSU_{D]}{}^{H]}(x)\delta^{(3)}(x-y)
\nn\\
\left\{\cpiP_{ABCD}(x),\cSU_{E}{}^{F}(y)\right\}_p&=&4\left(\delta_E^G\delta_{[A}^{F}-\frac{1}{8}\delta^F_E\delta_{[A}^G\right)\cpiP_{BCD]G}(x)\delta^{(3)}(x-y)
\nn\\
\left\{\cSU_{A}{}^{B}(x),\cSU_{E}{}^{F}(y)\right\}_p&=&\left(\delta_{E}^{B}\cSU_A{}^F -\delta_{A}^{F}\cSU_E{}^B\right)\delta^{(3)}(x-y).
\ee
Note that we obtain the structure functions of the Lie algebra $\mathfrak{e}_{7(7)}$ (see also eq. (4.1.47) in \cite{H08}) on the r.h.s. of the last three equations, which was expected because the $E_{7(7)}$-symmetry has been kept manifest throughout the construction. Thus, we have completed the first step of constructing the scalar sector of phase space. Before continuing with the second one, we want to remark that our approach of taking the group element $\cV\in E_{7(7)}$ as an independent variable is of course completely equivalent to the one presented in \cite{MN94}, where the algebra element $h\in \mathfrak{e}_{7(7)}$ with $\cV=\exp(h)$ is taken as the fundamental coordinate.\\

The guiding principle in this article is to maintain the manifest $E_{7(7)}$-covariance in all expressions. This implies in particular that we dispense with fixing an explicit gauge for the $E_{7(7)}/(SU(8)/\Z_2)$ coset $\cV$. For the discussion of the scalar sector of phase space, we hence keep $133$ momenta provided by $\cpiP_{ABCD}$ and $\cSU_A{}^B$, but we will have to impose a weak $SU(8)$-constraint that will project on the physical constraint hypersurface with only $70$ independent momenta. Starting from the bosonic Lagrangian $\cL_{\text{bos}}$ (\ref{LaLL}), the relation of the new momenta (\ref{momenta}) to the variation of the Lagrangian $\delta\cL$ (\ref{Lvar}) leads with (\ref{vielbeinG},\,\ref{BA}) to the weak equations (in complete analogy to the weak constraint $\cPhi_\cMa^i\approx 0$ (\ref{constr}) for the vector fields):\footnote{We are using Dirac's notation ``$\approx$'' for weak equalities \cite{D64}, i.e. for equations that only hold on the constraint hypersurface $\cPhi=0$.}
\begin{subequations}\label{weakP}
\be\label{weakP1}
\cpiP_{ABCD}&\approx&-\frac{e_4}{12}(\cvA^0)_{ABCD}
\\
\cSU_A{}^B&\approx&0.
\label{weakP2}
\ee
\end{subequations}
Since we are relating the seventy scalar velocities $\dot{\cV}$ to $133$ momenta, the identities (\ref{weakP}) only hold in the weak sense \cite{D64}, i.e. after imposing the $SU(8)$-constraint $\cSU_A{}^B\approx0$. We also want to remark that we could have dispensed with the $SU(8)$-constraint completely, if we had fixed a particular gauge for the coset $\cV\in E_{7(7)}/(SU(8)/\Z_2)$: Then, we would have only  had seventy independent momenta in the variation $\delta \cL$ (\ref{Lvar}) and consequently, the ``momenta'' $\cpiP_{ABCD}$ and $\cSU_A{}^B$ (\ref{momenta},\,\ref{Lvar}) would have been (non-vanishing!) functions of those. Their Poisson brackets, however, would not exhibit the $E_{7(7)}$-symmetry in a manifest way any more, in contradistinction to (\ref{Poisson3}). Since we want to keep the $E_{7(7)}$-covariance manifest, we hence have to rely on the parametrization of phase space in terms of $133$ momenta, subject to the $SU(8)$-constraint $\cSU_A{}^B\approx0$.\\

Before stating the Hamiltonian of the scalar sector, we will proceed with some comments on the Hamiltonian description of gravity. The conjugate momentum $\cpig^{\mu\nu}$ of the metric $g_{\mu\nu}$ is defined in the standard way
\be\label{momdefi2a}
\cpig^{\mu\nu}&:=&\frac{\p \cL}{\p \dot{g}_{\mu\nu}}
\ee
and we have to generalize the Poisson bracket defined for the vector system in (\ref{Poisson}) and for the scalars in (\ref{Poisson3}) to also yield the canonical equal time relation
\be\label{Poisson2}
\left\{g_{\mu\nu}(x),\cpig^{\sigma\tau}(y)\right\}_p&=&\delta_\mu^{(\sigma}\delta_\nu^{\tau)}\delta^{(3)}(x-y).
\ee
It is a standard exercise to compute the Hamiltonian of the Einstein--Hilbert Lagrangian \cite{ADM62}. Before stating the result, we want to emphasize some characteristic features of this computation that will be of crucial importance for our purpose. The first observation is that the Legendre transformation of the Einstein--Hilbert Lagrangian is singular as for the case of the vectors in the preceding section. The gravity system is constrained by the vanishing of the momenta $\cpig$ and $\cpig_i$ conjugate to the lapse $N$ and the shift $N^i$ (\ref{metric}). The only momenta among the $\cpig^{\mu\nu}$ (\ref{momdefi2a}) that appear in the Hamiltonian after imposing the constraints, are the momenta $\cpig^{ij}$ of the spatial metric $g_{ij}=\h_{ij}$ (\ref{metric}). Hence, in principle, we would have to repeat the analysis of the preceding section to find the total Hamiltonian that preserves all the constraints of the combined system. However, we can use a short-cut, which will greatly simplify our analysis. This is due to the following observation: the constraints of the combined metric, scalar and vector system are {\it decoupled} from each other. The decoupling is due to the facts that the constraints $\cPhi^i_\cMa$ (\ref{constr}) for the vector part of the system are independent of both the metric and the scalar fields as well as their momenta, that the $SU(8)$-constraint $\cSU_A{}^B$ (\ref{weakP2}) only restricts the scalar degrees of freedom and that the pure gravity constraints $\cpig\approx0$, $\cpig_i\approx0$, $\cH^{\text{grav}}\approx0$ and $\cH_k^{\text{grav}}\approx0$ \cite{ADM62} only depend on gravity variables, of course. Hence, we can conclude that the total Hamiltonian $H_{\text{tot}}$ of the combined system is determined by adding the three (total) Hamiltonians, the one of vacuum general relativity, the one of the scalars and the vector Hamiltonian defined in (\ref{H1tot}). In particular, $H_{\text{tot}}$ is of the same shape as the total Hamiltonian of the vector system (\ref{H1tot}):
\be\label{H2tot}
H_{\text{tot}}
\,=\,\int d^3 x\, N \cH + N^k \cH_k +  \cA_0{}^\cMa \p_i\cPhi^i_\cMa +\lambda \cpig +\lambda^i \cpig_{i} + \lambda_B{}^A\cSU_A{}^B.
\ee
Apart from the Lagrange multiplier $\cA_0{}^\cMa$ from section \ref{vecHam}, we have to introduce further Lagrange multiplier fields $\lambda$, $\lambda^i$ and $\lambda_B{}^A$ enforcing the primary constraints $\cpig=0$, $\cpig_{i}=0$ and $\cSU_A{}^B=0$.\footnote{A good review of this topic is \cite{T01}.} The complete energy and momentum densities $\cH$ and $\cH_k$ in the total Hamiltonian $H_{\text{tot}}$ (\ref{H2tot}) contain the contributions $\cH^{\text{vec}}$ and $\cH^{\text{vec}}_k$ (\ref{dense2}) (from the Lagrangian $\cL_{\text{vec}}$) as well as both the scalar and the pure gravity contributions $\cH^{\text{scal}},\cH^{\text{scal}}_k$ and $\cH^{\text{grav}},\cH^{\text{grav}}_k$ respectively:
\begin{subequations}\label{dense3}
\be
\cH&:=&\cH^{\text{grav}}+\cH^{\text{vec}} +\cH^{\text{scal}}\\
\cH_k&:=&\cH^{\text{grav}}_k+\cH^{\text{vec}}_k +\cH^{\text{scal}}_k.
\ee
\end{subequations}
The scalar contribution to the energy and momentum densities reads with the identities (\ref{BA},\,\ref{weakP1}):
\begin{subequations}\label{dense4}
\be
\cH^{\text{scal}}&=&\frac{6}{e_3} \cpiP^{ABCD}\cpiP_{ABCD} +\frac{e_3}{24}\h^{ij}(\cvA_i)^{ABCD}(\cvA_j)_{ABCD}
\\
\cH^{\text{scal}}_k&=&\cpiP^{ABCD}(\cvA_k)_{ABCD}.
\ee
\end{subequations}
Since the Poisson bracket of any phase space variable with the $SU(8)$-constraint $\cSU_A{}^B$ is equivalent to its transformation under an $\mathfrak{su}_8$-Lie algebra action (\ref{Poisson3}), the Poisson brackets of $\cSU_A{}^B$ with the $SU(8)$-scalars $\cH$ and $\cH_k$ vanish strongly. This implies in particular that $\cSU_A{}^B$ is a first-class constraint, whose associated gauge transformations are local $SU(8)$-transformations.\\

Finally, we also state the pure gravity part of the Hamiltonian that takes the form \cite{ADM62}
\begin{subequations}\label{dense5}
\be
\cH^{\text{grav}}&=&\frac{1}{e_3}\left(\cpig^{ij}\cpig^{kl}\h_{ik}\h_{jl}-\frac12(\cpig^{ij}\h_{ij})^2\right)
 -e_3 R^{(\h)} 
\\
\cH^{\text{grav}}_k&=&-2\h_{ki}\nabla^{(\h)}_j\cpig^{ij}
\ee
\end{subequations}
with the spatial covariant derivative $\nabla^{(\h)}_jC^i=\p_j C^i + (\G^{(\h)})_{jk}^iC^k$ and the spatial curvature scalar $R^{(\h)}$ constructed from the spatial metric $\h_{ij}$ \cite{ADM62}. Furthermore, the definition of the momenta (\ref{momdefi2a}) leads to the relation
\be\label{momdefi3}
\cpig^{ij}&=& e_4\h^{ik}\h^{jl}[(\G^{(g)})^0_{kl} -h_{kl}(\G^{(g)})^0_{rs}h^{rs}].
\ee
Thus, we have collected all ingredients to check general covariance, what we shall do in the next section. 
\end{subsection}

\begin{subsection}{Dirac algebra and $\Diff(4)$-action on vector fields}\label{algebra2}
It is well-known \cite{HT88,D64} that the invariance of a theory under general coordinate reparametrization is equivalent to the requirement that the evolution must not depend on the path that links a given initial spacelike hypersurface to a given final one. This independence is guaranteed, if the energy and momentum density satifies the hypersurface deformation algebra \cite{D64}, which reads (for equal times and $x,y$ being coordinates on the spatial slice) on the constraint hypersurface:
\begin{subequations}\label{algebra}
\be
\left\{\cH(x),\cH(y)\right\}_p &\approx&\left(\h^{ij}(x)\cH_i(x) +\h^{ij}(y)\cH_i(y)\right)\frac{\p}{\p x^j}\delta^{(3)}(x-y)\\
\left\{\cH_i(x),\cH(y)\right\}_p &\approx&\cH(x)\frac{\p}{\p x^i}\delta^{(3)}(x-y)
\\
\left\{\cH_i(x),\cH_j(y)\right\}_p &\approx&\cH_i(y)\frac{\p}{\p x^j}\delta^{(3)}(x-y)
+\cH_j(x)\frac{\p}{\p x^i}\delta^{(3)}(x-y).
\ee
\end{subequations}
Enhanced with the Poisson algebra relations (\ref{Poisson}), (\ref{Poisson2}) and (\ref{Poisson3}), it is a straightforward, but tedious exercise to verify the relations (\ref{algebra}) for the complete energy and momentum densities $\cH$ and $\cH_k$ (\ref{dense3}). We have used the following short-cut for this computation: Due to the equivalence of the closure of the energy-momentum algebra (\ref{algebra}) to general covariance, it is guaranteed that it holds for the metric-scalar system, whose action (\ref{action2}) exhibits manifest $\Diff(4)$-invariance, before coupling it to the vectors. Therefore, it turns out to be sufficient to verify the following algebraic relations:
\beg
\left\{\cH^{\text{vec}}(x),\cH^{\text{vec}}(y)\right\}_p
 &=&\left(\h^{ij}(x)\cH^{\text{vec}}_i(x) +\h^{ij}(y)\cH^{\text{vec}}_i(y)\right)\frac{\p}{\p x^j}\delta^{(3)}(x-y)\nn\\
 \left\{\cH_i(x),\cH^{\text{vec}}(y)\right\}_p
 &=&
\cH^{\text{vec}}(x)\frac{\p}{\p x^i}\delta^{(3)}(x-y)
+\frac{\delta \cH^{\text{vec}}(y)}{\delta \cpi^i_\cMa(x)}\p_j\cPhi^j_\cMa(x)
\\
\left\{\cH^{\text{vec}}_i(x),\cH^{\text{vec}}_j(y)\right\}_p
 &=&
\cH^{\text{vec}}_i(y)\frac{\p}{\p x^j}\delta^{(3)}(x-y)
+\cH^{\text{vec}}_j(x)\frac{\p}{\p x^i}\delta^{(3)}(x-y)\nn\\
&&+\frac{\delta \cH^{\text{vec}}_j(y)}{\delta \cpi^i_\cMa(x)}\p_j\cPhi^j_\cMa(x).
\eeg
Here, we have used the fact that the vector part $\cH^{\text{vec}}_k$ (\ref{dense2}) of the momentum density does not depend on the metric nor on the scalars and hence, its Poisson brackets with $\cH^{\text{grav}}$ and $\cH^{\text{scal}}$ vanish trivially. Therefore, we only have to keep the complete momentum density $\cH_i$ on the l.h.s. of the second equation. Note that the appearance of the two terms proportional to the first-class constraints $\p_j\cPhi^j_\cMa$ in the relations $\{\cH_i(x),\cH(y)\}_p$ and $\{\cH_i(x),\cH_j(y)\}_p$ could have been expected due to the analogy to classical electrodynamics coupled to general relativity, in which case the algebra exactly has the same shape [the index $\cMa$ being trivial]. In this way, we have proved that the hypersurface deformation algebra or Dirac algebra \cite{D64} is fulfilled in our case. Thus, the complete bosonic action exhibits general covariance \cite{HT88}. The inclusion of the fermions is not expected to modify this behaviour.\\

The construction of the Hamiltonian provides us with a further insight, namely the action of an infinitesimal diffemorphism $\xi^\nu$ on the fields of our theory. The preservation of the two primary constraints of (vacuum) general relativity $\cpig\approx0$ and $\cpig_i\approx0$ in the evolution entails the vanishing of the energy and the momentum density $\cH\approx0$ and $\cH_k\approx0$ as secondary constraints. Since the Hamiltonian is first-class, these two constraints give rise to gauge transformations, which turn out to be space-time diffeomorphisms \cite{T01}. In complete analogy to the gauge transformations that correspond to the first-class constraints $\p_j\cPhi^j_\cNa$ (\ref{gaugeTrafo}), the diffeomorphism action on any function $f$ on phase space is hence generated by the Poisson bracket, i.e.\footnote{The transformation of the vector field components $\xi^\nu$ into $\xi^\bot$ and $\tilde{\xi}^j$ follows the standard decomposition of a vector into components normal and perpendicular to the spatial slices. Starting from $n_\nu dx^\nu=-Ndx^0$, we obtain the vector $n^\mu:=g^{\mu\nu}n_\nu$, i.e. $n^\mu\p_\mu = \frac{1}{N}\p_0 -\frac{N^j}{N}\p_j$ with (\ref{metric}). The identity $\xi^\nu\p_\nu=\xi^\bot n^\nu\p_\nu +\tilde{\xi}^j\p_j$ then fixes the relations between $\xi^\mu$ and $(\xi^\bot,\tilde{\xi}^k)$ in (\ref{diffeo1}). Further details can be found in \cite{PSS96}.}
\be\label{diffeo1}
\delta_\xi f(x)&:=&-\int d^3y\, \{\xi^\bot(y)\cH(y) +\tilde{\xi}^k(y)\cH_k(y),f(x)\}_p\\
\text{with}\quad \xi^\bot&:=&N\xi^0\quad\text{and}\quad \tilde{\xi}^k\,\,:=\,\,\xi^k +\xi^0N^k.
\nn
\ee
For the (spatial) metric $g_{ij}=\h_{ij}$ and the scalars $G_{\cMa\cNa}$, this formula reproduces the standard $\Diff(4)$-actions (where we use the expression (\ref{momdefi3}) for $\cpig^{ij}$):
\begin{subequations}\label{trafoG}
\be
\delta_\xi g_{ij}&=& \xi^\mu\p_\mu g_{ij} +2g_{\mu (i}\p_{j)}\xi^\mu\\
\delta_\xi G_{\cMa\cNa}&=& \xi^\mu\p_\mu G_{\cMa\cNa}
.
\ee
\end{subequations}
For the vector potential $\cA_i{}^\cMa$, we obtain on the (primary) constraint hypersurface $\cPhi=0$:
\be\label{diffeo2}
\delta_\xi \cA_i{}^\cMa&=& \xi^0 \cB_i{}^\cMa + 2\tilde{\xi}^k\p_{[k}\cA_{i]}{}^\cMa\\
&=& \xi^\mu \p_\mu \cA_i{}^\cMa +\cA_\mu{}^\cMa \p_i\xi^\mu 
-\p_i\left(\xi^\mu \cA_\mu{}^\cMa\right)
+\xi^0\left(\cB_i{}^\cMa -\cE_i{}^\cMa\right).
\nn
\ee
The important observation is that this definition of a general coordinate transformation only agrees with the standard transformation of a vector field after imposing the equation of motion $\cE_i{}^\cNa  = \cB_i{}^\cNa$ (\ref{BWGL5d}) and upon adding a gauge transformation $\delta_g \cA_i{}^\cMa=\p_i(\xi^\mu \cA_\mu{}^\cMa)$.\footnote{The entanglement of the $\Diff(4)$-action with a gauge transformation in Einstein-Yang-Mills theories has already been observed in \cite{PSS99}.} What is more, the spatial components $\cA_i{}^\cMa$ of all the $56$ vector potentials transform under diffeomorphisms in a local way.\\

Finally, we can check the invariance of the action $S_{\text{vec}}$ (\ref{action}) under general coordinate transformations. With the standard extension of the metric transformation $\delta_\xi g$ (\ref{trafoG}) to also include the lapse $N$ and the shift $N^k$, i.e. $\delta_\xi g_{\nu\rho}= \xi^\mu\p_\mu g_{\nu\rho} +2g_{\mu (\nu}\p_{\rho)}\xi^\mu$, we have verified the relation
\be
\delta_\xi S_{\text{vec}}&=&0.
\ee
Due to the manifest $\Diff(4)$-invariance in the other parts of the bosonic action $S_{\text{bos}}$ (\ref{action2}), we can therefore conclude $\delta_\xi S_{\text{bos}}=0$. Thus, we have shown that the theory is invariant under general coordinate transformations. We want to emphasize again that we have to use the diffeomorphism action $\delta_\xi\cA_i{}^\cMa$ in the form (\ref{diffeo2}) in order to prove the $\Diff(4)$-invariance of the action. This is not equivalent to the standard form as far as this computation is concerned, because one must not use the equations of motion within the action. In other words, we had to use the diffeomorphism action on the vector potentials $\cA_i{}^\cMa$ prescribed by the Hamiltonian formalism in order to guarantee the general covariance of the theory. In contradistinction to a theory with manifest general coordinate invariance, this action differs from the standard vector-field transformation {\it off-shell}. This fact has already been observed by Henneaux and Teitelboim in \cite{HT88,HT87}.\\

The inclusion of the fermions is not expected to lead to any complications. First, one has to extend the bifermionic contribution to the Hamiltonian involving the vector field beyond the constraint hypersurface $\cPhi=0$. This has to be done again in such a way that the constraints $\cPhi^i_\cMa$ (\ref{constr},\,\ref{bedingung}) are conserved. Then, one can completely follow the analysis used for the bosonic case to prove the general covariance of the complete theory.\footnote{Of course, we have to switch to the `vielbein frame' for both gravity and the scalar fields (as explained in section \ref{vielbeinS}) in order to couple the bosons to fermions. Note in particular that the transformation of the vector field $\cA_i{}^\cMa$ under a general coordinate transformation (\ref{diffeo2}) will be modified by the bifermionic contribution $T_k{}^\cNa$ (\ref{BWGL6c}) such that its on-shell equivalence to the standard form still holds.} 
\end{subsection}

\begin{subsection}{Conserved Noether charge}\label{CNoet}
To complete our analysis of the $E_{7(7)}$-covariant formulation of $d=4$ $\cN=8$ supergravity in the Hamiltonian formalism, we want to address the Noether charge $Q$. This is defined from the conserved current $j^\mu$ (\ref{Noether7}) in the usual way by $Q=\int d^3x\,j^0$ (\ref{chargeE}). As before, we want to focus on the purely bosonic part in this section, whose form can be taken from eq. (\ref{Noether2}) after substituting $\cpiP_{ABCD}$ (\ref{weakP1}):
\be\label{Q1}
Q&=& \int d^3x\,\Big(\frac12
\big(\cV_\cPa{}^{AB}\cV^{CD,\cNa} +\frac{1}{4!}\e^{ABCDEFGH}\cV_{\cPa,EF}\cV_{GH}{}^\cNa\big)\cpiP_{ABCD}
\nn\\
&&
+\frac{1}{16}\e^{ijk}\cF_{ij}{}^\cMa\Omega_{\cPa\cMa}\cA_k{}^\cNa
+F_{i\,\cPa}{}^{\cNa\cMa}(\cA,\cpi)\cPhi^i_\cMa\Big)
\Lambda_\cNa{}^\cPa.
\ee
Due to the singular nature of the Legendre transformation, the charge is only well-defined on the (primary) constraint hypersurface $\cPhi=0$ a priori. In analogy to the construction of the total Hamiltonian (\ref{Htotal}), we had to include a function $F_{i\,\cPa}{}^{\cNa\cMa}(\cA,\cpi)$ in the expression for the conserved charge $Q$ (\ref{Q1}) that is fixed by the requirement that $Q$ preserve the (primary) constraint hypersurface $\cPhi=0$. In other words, $F$ is determined by the equation
\be\label{QP1}
\{Q,\cPhi^i_\cMa(x)\}_p&\approx&0
\ee
that only has to hold on the hypersurface $\cPhi=0$. Its general solution is provided by the sum of a special solution and the solution of the associated homogeneous problem as in the construction of the total Hamiltonian in eq. (\ref{homog}). Thus, we arrive at the simpler form
\be\label{Q2}
Q&=& \int d^3x\,
\Big(\frac12
\big(\cV_\cPa{}^{AB}\cV^{CD,\cNa} +\frac{1}{4!}\e^{ABCDEFGH}\cV_{\cPa,EF}\cV_{GH}{}^\cNa\big)\cpiP_{ABCD}
\nn\\
&&-\cpi^k_\cPa\cA_k{}^\cNa
+f_{\cPa}{}^{\cNa\cMa}\p_i\cPhi^i_\cMa\Big)
\Lambda_\cNa{}^\cPa,
\ee
where $f_{\cPa}{}^{\cNa\cMa}$ is an arbitrary function of space-time. $Q$ then satisfies the expected extension of eq. (\ref{QP1}) beyond the constraint hypersurface:
\be\label{QP2}
\{Q,\cPhi^i_\cMa(x)\}_p&=&-\Lambda_\cMa{}^\cPa \cPhi^i_\cPa(x).
\ee
The last term on the r.h.s of eq. (\ref{Q2}) is a gauge transformation $\delta_g$ generated by the first-class constraints $\p_i\cPhi^i_\cMa$ (\ref{gaugeTrafo}). Since $\p_i\cPhi^i_\cMa$ has vanishing Poisson bracket with both the Hamiltonian (\ref{bedingung}) and with itself (\ref{homog}), we can without loss of generality drop this term in the charge $Q$ (\ref{Q2}) by setting $f_{\cPa}{}^{\cNa\cMa}=0$. \\

For the computation of Poisson brackets of $Q$ with the energy and momentum densities $\cH$ and $\cH_k$ (\ref{dense3}), it is crucial to recall that the constant matrix $\Lambda_\cNa{}^\cPa$ is $\mathfrak{e}_{7(7)}$-valued. Since this is equivalent to $\cV^{-1}\Lambda\cV$ being $\mathfrak{e}_{7(7)}$-valued, we can without loss of generality define in analogy to the Maurer--Cartan form (\ref{BA}):
\begin{subequations}\label{lambdaflach}
\be
\Lambda^{ABCD}&:=&\cV^{CD,\cNa}\Lambda_\cNa{}^\cPa\cV_\cPa{}^{AB}
\\
2\Lambda_{[A}{}^{[C}\delta_{B]}^{D]}&:=&\cV_{AB}{}^{\cNa}\Lambda_\cNa{}^\cPa\cV_\cPa{}^{CD}.
\ee
\end{subequations}
The (non-constant) coefficients $\Lambda^{ABCD}$ and $\Lambda_{A}{}^{C}$ are subject to the same constraints as $\cvA$ and $\cvB$ in (\ref{BA},\,\ref{BA2}).\footnote{Note that we could have equivalently imposed constraints on the components of $\Lambda$ instead. We have chosen to restrict the `dressed' version $\cV^{-1}\Lambda\cV$ (\ref{lambdaflach}), because thus, we do not have to split the summation of the $56$ `curved' indices $\cNa,\cPa$ and hence, the $E_{7(7)}$-covariance remains manifest.} Using these abbreviations, the Noether charge $Q$ (\ref{Q2}) takes the form:
\be\label{Q3}
Q&=&\int d^3x
\Big(\Lambda^{ABCD}\cpiP_{ABCD}
-\cpi^k_\cPa\cA_k{}^\cNa \Lambda_\cNa{}^\cPa
\Big).
\ee
Keeping in mind the restrictions on the $\cV$-dependent coefficients $\Lambda^{ABCD}$ and $\Lambda_{A}{}^{C}$ (\ref{BA2},\,\ref{lambdaflach}), it is a straightforward exercise using the Poisson brackets (\ref{Poisson},\,\ref{Poisson3}) to verify that the Noether charge $Q$ (\ref{Q3}) commutes with the energy and momentum densities $\cH$ and $\cH_k$ (\ref{dense3}) on the constraint hypersurface:
\begin{subequations}\label{QH}
\be
\{Q,\cH(x)\}_p&=&8\Lambda^{ABCE}(x)\cpiP_{ABCD}(x)\cSU_E{}^D(x)\\
\{Q,\cH_k(x)\}_p&=&\frac23\Lambda^{ABCE}(x)(\cvA_k)_{ABCD}(x)\cSU_E{}^D(x).
\ee
\end{subequations}
Furthermore, for any two parameters $\Lambda_1,\Lambda_2\in \mathfrak{e}_{7(7)}$, the Poisson bracket of the associated charges $Q_{\Lambda_1},Q_{\Lambda_2}$ reproduces the Lie algebra $\mathfrak{e}_{7(7)}$ on the constraint hypersurface:
\be\label{Qalge}
\left\{Q_{\Lambda_1},Q_{\Lambda_2}\right\}_p&=& Q_{[\Lambda_1,\Lambda_2]} +\frac23\Lambda_1^{ABCD}\Lambda_{2,ABCE}\cSU_D{}^E
\ee
with $[\Lambda_1,\Lambda_2]_\cNa{}^\cPa=\Lambda_{1,\cNa}{}^\cMa \Lambda_{2,\cMa}{}^\cPa -\Lambda_{2,\cNa}{}^\cMa \Lambda_{1,\cMa}{}^\cPa$. Together with a corollary of relation (\ref{QP2}) being 
\be\label{QP3}
\{Q,\p_i\cPhi^i_\cMa(x)\}_p&=&-\Lambda_\cMa{}^\cPa \p_i\cPhi^i_\cPa(x),
\ee
the Poisson bracket of $Q$ with all the first-class constraints is first-class and hence, the charge $Q$ is first-class itself \cite{D64} (as expected). In analogy to the gauge transformations in (\ref{gaugeTrafo}), the global $\mathfrak{e}_{7(7)}$ transformations are generated by taking Poisson brackets of $Q$ with any function on phase space. For the scalars $G_{\cMa\cNa}$ and the vectors $\cA_i{}^\cMa$, we obtain in particular [keeping in mind (\ref{lambdaflach})]
\begin{subequations}\label{Qaction}
\be
\delta_\Lambda G_{\cMa\cNa}&:=&\{Q,G_{\cMa\cNa}\}_p
\,\,=\,\,-2\Lambda_{(\cMa}{}^\cPa G_{\cNa)\cPa}\\
\delta_\Lambda \cA_i{}^\cMa&:=&\{Q,\cA_i{}^\cMa\}_p
\,\,=\,\,\Lambda_{\cNa}{}^\cMa \cA_i{}^\cNa.
\ee
\end{subequations}
To summarize our findings, the Noether charge $Q$ (\ref{Q3}) of the $E_{7(7)}$ transformations in $d=4$ $\cN=8$ supergravity shows all the properties as the one of an ordinary global symmetry group. Apart from being conserved (\ref{QH}), $Q$ generates infintesimal $E_{7(7)}$ transformations (\ref{Qaction}) (satisfying $[\delta_{\Lambda_1},\delta_{\Lambda_2}]\approx\delta_{[\Lambda_1,\Lambda_2]}$) and $Q$ is invariant under the gauge transformations $\delta_g$ (\ref{gaugeTrafo}), which follows from the relation (\ref{QP3}).
\end{subsection}

\begin{subsection}{Quantization and Dirac brackets}\label{Dirac2}
In this section, we shall briefly comment on the particularities the second-class constraints among $\cPhi^i_\cMa=0$ (\ref{constr}) impose on the canonical quantization procedure. Following Dirac's standard analysis \cite{D64}, we have to solve for these constraints by replacing the Poisson brackets by Dirac brackets prior to quantization. Since the treatment of the pure gravity and the scalar part is fairly standard, we will focus on the vector fields $\cA_i{}^\cMa$. We emphasize that it is consistent to separate the discussion of the constraints $\cPhi^i_\cMa$ of the vector fields from the others, because the former exclusively depend on the vector field degrees of freedom $(\cA_i{}^\cMa,\cpi_\cNa^j)$. In order to define Dirac brackets for $\cA_i{}^\cMa$ and their conjugate momenta $\cpi_\cNa^j$, we have to invert the ``matrix'' $C$ defined by the Poisson bracket of the second-class constraints with each other:
\be\label{CDefi}
C^{ij}_{\cMa\cNa}(x,y)&:=&\left.\left\{\cPhi^i_\cMa(x),\cPhi^j_\cNa(y)\right\}_p\right|_{\text{second-class}}
\ee
The Dirac bracket of any two functions $X,Y$ of $\cA_i{}^\cMa$ and $\cpi_\cNa^j$ is then defined in the standard way \cite{D64,HT87}:
\be\label{Dirac}
\left\{X(x),Y(y)\right\}_D &:=&\left\{X(x),Y(y)\right\}_p \\
&&\!\!\!\!\!\!-\int d^3v d^3 w 
\left\{X(x),\cPhi^i_\cMa(v)\right\}_p (C^{-1})_{ij}^{\cMa\cNa}(v,w) \left\{\cPhi^j_\cNa(w),Y(y)\right\}_p.
\nn
\ee
At a first glance, it looks as if we had arrived at a dead-end, because we cannot separate the second-class constraints from the first-class ones within $\cPhi^i_\cMa(x)=0$ (\ref{constr}). However, it will not be necessary for the quantization of our system to perform this step explicitly as we shall show next. In order to examine the implications of the Dirac procedure, let us assume for a while that we can eliminate the second-class constraints from our system in the standard way. This would imply that we could then without loss of generality impose these constraints also beyond the constraint hypersurface, since the Dirac bracket of any function $\xi$ of phase space with the second-class constraints vanishes by construction \cite{D64}. In particular, we could substitute the vector part of the Hamiltonian densities $\cH^{\text{vec}}$ and $\cH^{\text{vec}}_k$ (\ref{dense2}) by the expressions
\begin{subequations}\label{denseD}
\be
\cH^{\text{vec}}&=&\frac{8}{e_3}h_{ij}G^{\cMa\cNa}\cpi^i_\cMa \cpi^j_\cNa
\\
\cH^{\text{vec}}_k&=&-8\e_{kij}\Omega^{\cMa\cNa}\cpi^i_\cMa \cpi^j_\cNa.
\ee
\end{subequations}
The important observation is that the phase space variables $\cA_i{}^\cMa$ have completely disappeared from the Hamiltonian densities. This statement trivially extends to the total Hamiltonian $H_{\text{tot}}$ (\ref{H2tot}) of the entire system, because the first class constraints $\p_i\cPhi^i_\cMa$ do not depend on them either. Therefore, we {\it do not need} to know the explicit expression for the Dirac bracket $\{\cA_i{}^\cMa(x),\cA_j{}^\cNa(y)\}_D$ in order to describe the evolution of the system. It is sufficient to know $\{\cA_i{}^\cMa(x),\cpi^j_\cNa(y)\}_D$ and $\{\cpi^i{}_\cMa(x),\cpi^j_\cNa(y)\}_D$, and these can be computed explicitly with the following identity of Poisson brackets:
\be\label{ident3}
\left\{\cPhi^i_\cMa(x),\cpi^j_\cNa(y)\right\}_p 
&=&
\frac12\left\{\cPhi^i_\cMa(x),\cPhi^j_\cNa(y)\right\}_p.
\ee
Therefore, we do not have to compute the inverse of $C$ (\ref{CDefi}) in a closed form in order to obtain the Dirac brackets that are necessary for the description of our system. Hence, we can in fact perform the Dirac procedure. The relevant Dirac brackets can then be obtained from their definition (\ref{Dirac}) with the standard Poisson bracket (\ref{Poisson}) and the constraints $\cPhi^i_\cMa$ (\ref{constr}):
\begin{subequations}\label{Dbracket}
\be
\left\{\cA_i{}^\cMa(x),\cpi^j_\cNa(y)\right\}_D 
&=&\frac12 \delta_i^j\delta^\cMa_\cNa\delta^{(3)}(x-y)
\\
\left\{\cpi^i_\cMa(x),\cpi^j_\cNa(y)\right\}_D
&=&\frac{1}{16}\e^{ijk}\Omega_{\cMa\cNa}\frac{\p}{\p x^k}\delta^{(3)}(x-y)
\ee
\end{subequations}
It is no surprise that the Dirac algebra of section \ref{algebra2} can also be reproduced using these Dirac brackets and the Hamiltonian densities in the form (\ref{denseD}). Note that this is the formulation that has been chosen in the Henneaux's and Teitelboim's article on ``chiral $p$-forms'' \cite{HT88}. For the classical analysis, the Dirac procedure is completely equivalent to the canonical description. However, in order to quantize the theory, it is important to elevate the Dirac brackets (\ref{Dbracket}) [and not the Poisson brackets (\ref{Poisson})] to commutators of operators in order to avoid inconsistencies \cite{D64}. Furthermore, let us remark that the treatment of the gauge invariance $\delta_g$ (\ref{gaugeTrafo}) should not involve any particular complications: it must be handled according to the usual Faddeev-Popov or BRST methods \cite{HT88}. \\

Before concluding, we want to briefly comment on the implications of the elimination of the second-class constraints on phase space. The constraint $\cPhi^i_\cMa\approx0$ (\ref{constr}) implies that the curl of the vector field $\cA$ can be identified with the associated momentum $\cpi$. Due to the gauge arbitrariness of $\cA$ however, the complete information of the vector fields $\cA$ is encoded in their momenta $\cpi$. Therefore, the elimination of the second-class constraints effectively {\it halves} this part of the phase space by eliminating the variables $\cA_i{}^\cMa$, which strongly reminds of the quantization of fermions.\newpage

The standard formulation of the phase space of maximal supergravity (which {\it violates} the $E_{7(7)}$-symmetry) can also be linked to this formulation quite easily. Given $28$ vector fields $\cA_i$ and their conjugate momenta $\cpi^i$ (being subject to the constraint $\p_i\cpi^i\approx0$), it is straightforward to relate the latter to the so-called $28$ dual potentials by the constraint $\cPhi^i_\cMa\approx0$ (\ref{constr}) because of the canonical form of the symplectic form $\Omega=\binom{\,\,0\,\,\,\,1}{-1\,\,0}$ \cite{H08}.\\

In other words, the $E_{7(7)}$-symmetry acts on the $56$ dimensional space built up from all vector fields $\cA$ and functions of their momenta $\cpi$ in any formulation of maximal supergravity. In particular, the dual potentials are no extraneous new objects, they have to be thought of as functions of the momenta of the ordinary $28$ vector fields. A further investigation of the implications for the quantization of maximal supergravity along these lines is beyond the scope of this article, but it is strongly expected that this approach will contribute to a better understanding of the quantization of maximal supergravity in four dimensions.
\end{subsection}
\end{section}

\begin{section}{Conclusion and outlook}
In this article, we have constructed a Lagrangian of $d=4$ $\cN=8$ supergravity that exhibits manifest $E_{7(7)}$-invariance off-shell without the necessity of introducing a Lagrange multiplier as in \cite{CJ79}. The key ingredient was that we dispensed with the usual form of manifest $\Diff(4)$-covariance for the terms in the Lagrangian density involving the $56$ vector fields. Nonetheless, the corresponding action functional $S$ is invariant under general coordinate transformations
\beg
\delta_\xi S&=&0.
\eeg
We have proved explicitly for the bosonic sector that this `hidden' form of $\Diff(4)$-covariance is manifest in the standard energy-momentum algebra \cite{D64} and that the Hamiltonian formulation of the theory shows exactly the same form as the one of a manifestly covariant theory (like electrodynamics). This analogy extends to the gauge transformations of all the $56$ vector fields. We have also computed the conserved $E_{7(7)}$ Noether charge $Q$ from first principles, which shows exactly the same properties as for the case of an ordinary global symmetry. \\

In our $E_{7(7)}$-invariant formulation of supergravity, we have furthermore verified that the action functional is invariant under supersymmetry transformations and that the supersymmetry algebra closes on the bosons. We have also shown explicitly that the equations of motion deduced from the $E_{7(7)}$-invariant Lagrangian agree with the ones of the standard formulation of $d=4$ $\cN=8$ supergravity. To establish the contact to maximal supergravity in its conventional form (i.e. as a Kaluza--Klein reduction on a seven torus of $D=11$ supergravity \cite{CJS78} that contains only $28$ vector potentials), it is necessary to break the manifest $E_{7(7)}$ symmetry by eliminating $28$ field strengths from the theory as shown in the Appendix. The crucial point is that this procedure does not affect the on-shell field content of the theory due to the nature of the twisted self-duality equation of motion for the $56$ vector fields (\ref{ImRe},\,\ref{selfdualC3}).\\

Concerning the issue of quantization, we have shown that the $E_{7(7)}$ invariant system is subject to second-class constraints that reduce the dimension of phase space. In view of the on-shell equivalence to $d=4$ $\cN=8$ supergravity (a theory with $28$ vector fields without secondary constraints), this result did not come as a surprise. However, the present analysis offers the possibility of keeping the $E_{7(7)}$ symmetry manifest thoughout the quantization procedure which may turn out to be an important tool for an improved understanding of the quantization of maximal supergravity. In particular, our formulation reveals the precise way the $E_{7(7)}$ symmetry acts on the phase space of $d=4$ $\cN=8$ supergravity for any formulation of the theory.\\

Furthermore, the possible UV-finiteness of $\cN=8$ supergravity as a quantum field theory \cite{Finite} has been conjectured to be linked to the $E_{7(7)}$ symmetry \cite{FiniteE7}. The present off-shell formulation of the $E_{7(7)}$-symmetric theory may serve as a tool to check these conjectures and to eventually decide whether maximal supergravity is finite as a quantum field theory or not. As a starting point for such an investigation, it looks promising to follow and to possibly extend the analysis of Kallosh and Kugo \cite{KK08} that was aimed at establishing a direct link between the computation of scattering amplitudes on the one hand and the Noether current of the $E_{7(7)}$-symmetry on the other hand. Furthermore, it would be interesting to investigate whether it is possible to construct fully $E_{7(7)}$-invariant higher curvature corrections, keeping in mind that the proposals using superfield techniques \cite{K81} do not match the field content correctly a priori: In view of their manifest $SU(8)$-covariance, one has to deal with $56$ independent vector fields in $d=4$ that have to be restricted by an additional (twisted self-duality) constraint as in the formulation of \cite{CJLP97}. For $d=4$ $\cN=8$ supergravity, this constraint has been shown to be expressible as an equation of motion, but it is unclear whether this possibility persists for higher curvature corrections to the theory. \\

Our complementary  formulation of maximal supergravity in a manifestly $E_{7(7)}$ covariant way may also open the door to a better understanding of duality symmetries of supergravity in general. Even for the classical theory, some questions still remain open. Apart from relating the present analysis to the light-cone formulation of $E_{7(7)}$ in \cite{BKR08}, it looks promising to investigate whether addressing the classification of gauged supergravities \cite{dWST02,dWST07} from this point of view leads to new insights.\\

Finally, we want to remark that this formulation of supergravity fulfills all the requirements to be compatible with the conjectures for extended symmetry groups of M-theory, notably $E_{10(10)}$ \cite{DHN02} and $E_{11(11)}$ \cite{W01}. In particular, it naturally connects to the analysis in \cite{H09}, where it was shown that $D=11$ supergravity can partially be derived from a reduction \`a la Kaluza--Klein from a $D=4+56$ dimensional exceptional geometry. This geometry is restricted in such a way that the original $D=4+56$ dimensional theory exhibits only an $\Diff(4)\times E_{7(7)}$-covariance, but no $\Diff(11)$-covariance. The latter can only be expected to appear as a ``hidden symmetry'' in the reduction to eleven dimensions. Our present formulation of maximal $d=4$ supergravity with off-shell $E_{7(7)}$-symmetry was in fact a missing ingredient for the completion of the proof that $D=11$ supergravity arises from this $4+56$ dimensional exceptional geometry by a simple Kaluza--Klein reduction on a $49$-torus as explained in \cite{H09}. Therefore, we naturally expect that the combination of the present results with the ones of \cite{H09} will also shed some new light on the $E_{10}$ and $E_{11}$ conjectures \cite{DHN02,W01}. These are particular cases of the most interesting interplay between the exceptional symmetries, supersymmetry and general covariance, of which our understanding is still inadequate and which promises many valuable new insights into the structures of supergravity.
\mbox{}\\

{\bf Acknowledgements}\\
I would like to thank Marc Henneaux for pointing out the reference \cite{HT88} and I am grateful to Thibault Damour for clarifying comments and valuable advice.
\end{section}

\renewcommand{\thesection}{A}
  \renewcommand{\theequation}{A.\arabic{equation}}
  \setcounter{equation}{0}
\begin{section}*{Appendix: Relation to $D=11$ supergravity}\label{App}
We use the same conventions as in \cite{H09}. In particular, our signature in $D=11$ is mostly plus $(-+\cdots +)$ and we normalize the representation matrices $\tilde{\G}^A\in \R^{32\times 32}$ of the $D=11$ Clifford algebra $\{\tilde{\G}^A,\tilde{\G}^B\}=2\eta^{AB}$ by $\G^{A_0\dots A_{10}}=\e^{A_0\dots A_{10}}\id_{32}$ with $\e^{0\,1\dots 10}=+1$. As in the main text, indices from the middle of the alphabet $M,N$ will denote curved or coordinate indices and the ones from its beginning $A,B=0,\dots,10$ will dress flat objects, i.e. whose curved indices have been contracted with the elfbein $E_M{}^A$. In this convention, the action of $D=11$ supergravity to leading order in fermions takes the form
\be\label{Actionneu}
	S &=& \int d^{11}x\det(E)\left(\frac{1}{4}\tilde{R} -\frac{1}{2}\check{\psi}_B\tilde{\G}^{BCD}\nabla_C\psi_D - \frac{1}{48}F_{B_1\dots B_4}F^{B_1\dots B_4}\nn
	\right.\\
	&& -\frac{1}{96}\left( \check{\psi}_{B_5}\tilde{\G}^{B_1\dots B_6}\psi_{B_6} +12\check{\psi}^{B_1}\tilde{\G}^{B_2B_3}\psi^{B_4}\right)F_{B_1\dots B_4} \nn\\
	&&\left. +\frac{2}{12^4}\epsilon^{B_1\dots B_{11}}F_{B_1\dots B_4}F_{B_5\dots B_8}A_{B_9\dots B_{11}}\right).
\ee
The bosonic fields being the elfbein $E_M{}^A$ and the three-form $A_{MNP}$ are linked to the Majorana fermions $\psi_M$ by the following supersymmetry transformations\footnote{The Majorana conjugate $\check{\psi}_M$ of the anticommuting gravitino $\psi_M$ is defined by multiplying the transposed spinor $\psi_M$ by $i\G^0$ such that the action $S$ is real as in \cite{FN76}.}
\begin{subequations}\label{Trafo1neu}
\be
	\delta^{(11)} {E_M}^A &=& \check{\ep}\tilde{\G}^A\psi_M\\
	\delta^{(11)} \psi_M &=& \nabla_M\ep +\frac{1}{144}\left(\tilde{\G}{{}^{N_1\dots N_4}}_M-8\delta_M^{N_1}\tilde{\G}^{N_2\dots N_4}\right)\ep F_{N_1\dots N_4} \\
	\delta^{(11)} A_{N_1\dots N_3} &=& -\frac{3}{2}\check{\ep}\tilde{\G}_{[N_1N_2}\psi_{N_3]}
\ee
\end{subequations}
with the usual $D=11$ spin connection $\nabla$. In performing the Kaluza--Klein reduction from $D=11$ to $d=4$ on a flat seven-torus, we choose without loss of generality an upper triangular gauge for the elfbein and thus establish contact to the vierbein $e_\mu{}^\alpha$ (\ref{vielbein}) of $d=4$ supergravity:
\begin{subequations}\label{v114b}
\be
E_\nu{}^{\alpha}&=& \Delta^{-\frac12}e_\nu{}^{\alpha}\\
E_\nu{}^{a}&=& B_{\nu}{}^n e_n{}^{a}\\
E_n{}^{\alpha}&=& 0\\
E_n{}^{a}&=& e_n{}^{a}.
\ee
\end{subequations}
The indices have the range $\mu,\alpha=0,\dots,3$ and $n,a=4,\dots,10$ and we have used the abbreviation $\Delta:=\det(e_m{}^a)$ \cite{dWN86}. It is well-known that the field content of the resulting four dimensional theory consists of $28$ vector fields and $70$ scalars apart from the metric $g_{\mu\nu}$. The proof that the seventy scalars form a non-linear $\sigma$-model based on the coset space $E_{7(7)}/(SU(8)/\Z_2)$ is standard \cite{CJ79} and it will not be repeated here. Instead, we focus on the $28$ vectors, being the $7$ graviphotons $B_\nu{}^m$ and the $21$ vectors $A_{\nu mn}$, and link these to the $56$ vector fields $\cA_i{}^\cMa$ that we have used in the $E_{7(7)}$-invariant action $S$ (\ref{action},\,\ref{action2},\,\ref{actionferm}) of $d=4$ $\cN=8$ supergravity. With the inverse elfbein
\begin{subequations}\label{v114}
\be
E_\alpha{}^{\nu}&=& \Delta^{\frac12}e_\alpha{}^{\nu}\\
E_\alpha{}^{n}&=& -\Delta^{\frac12}e_\alpha{}^{\nu}B_{\nu}{}^n\\
E_a{}^{\nu}&=& 0\\
E_a{}^{n}&=& e_a{}^{n}
\ee
\end{subequations}
we switch to flat coordinates by defining
\begin{subequations}\label{vSym73d}
\be
\cA_\alpha{}^a 
&:=&
E_\alpha{}^\nu E_m{}^{a} B_\nu{}^m
\\
     \cA_{\alpha,cd} 
     &:=&
     -\sqrt{2}E_\alpha{}^\nu E_c{}^{m_1} E_d{}^{m_2} A_{\nu m_1m_2}.
\ee
\end{subequations}
These are then related to the $56$ vector fields in the vielbein frame $\cA_\alpha{}^{AB}:=e_\alpha{}^\mu \cV_\cMa{}^{AB}\cA_\mu{}^\cMa$ (\ref{flach1}) in the following way:
\be\label{ADec}
\cA_\alpha{}^{AB}
&=:&
\frac{1}{4i}{\G_a}^{AB}
\left(\cA_\alpha{}^{a}+i\eta^{ac}\cA_{\alpha,c}\right)
\\
&&
 +\frac{1}{4\sqrt{2}}{\G_{ab}}^{AB}
 \left(\cA_{\alpha}{}^{ab}+i\eta^{ac}\eta^{bd}\cA_{\alpha,cd}\right).
 \nn
\ee
Here, we are using purely imaginary $\G$-matrices satisfying the Euclidean Clifford algebra in $d=7$ $\{\G^a,\G^b\}=2\eta^{ab}$ with $\eta=\text{diag}(+++++++)$. We use the normalizations $\G^{a_1\dots a_7}=-i\e^{a_1\dots a_7}\id$ and $\e^{1\,2\,3\,4\,5\,6\,7}=+1$ \cite{dWN86,H09}. These relations enable us to relate $28$ field strengths among $\cF_{\alpha\beta}{}^{AB}$ (\ref{flach1}) to the four-form $F_{\alpha\beta cd}$ and to the graviphoton field strenghths $F_{\alpha\beta}{}^a$:
\begin{subequations}\label{vSym26d}
\be
F_{\alpha\beta}{}^a
&:=&
2E_{[\alpha}{}^\mu E_{\beta]}{}^\nu E_m{}^{a} \p_\mu B_\nu{}^m\,\,=\,\,2E_{[\alpha}{}^\mu E_{\beta]}{}^N  \p_\mu E_N{}^{a}
\nn\\
&=&
-\frac{i\Delta^{\frac12}}{2}\G^a{}_{AB}\Rea\left(\cF_{\alpha\beta}{}^{AB}\right)\\
     F_{\alpha\beta cd} 
     &:=&
     2E_{[\alpha}{}^\mu E_{\beta]}{}^{M_1} E_c{}^{M_2}E_d{}^{M_3} \p_\mu A_{M_1\dots M_3}
     \nn\\
          &=&
\frac{\Delta^{\frac12}}{4}\G_{cd}{}_{AB}\Ima\left(\cF_{\alpha\beta}{}^{AB}\right).
\ee
\end{subequations}
Note that the curved indices have the range $N,M_i=0,\dots,10$, whereas $A,B=1,\dots,8$ are $\G$-matrix indices. Furthermore, we want to emphasize that we had to associate the field strength $F_{\alpha\beta}{}^a$ to the real part of $\cF_{\alpha\beta}{}^{AB}$ and $F_{\alpha\beta cd}$ to its imaginary part (or vice versa), because both the field strengths and their corresponding potentials $B_\nu{}^m$ and $A_{\nu mn}$ cannot be combined into a common $Sl(8)\subset E_{7(7)}$ representation $\textbf{28}$. This is due to their different position of the curved indices $m,n=4,\dots,10$ indicating contragredient representations of $Gl(7)$. Note that this subtlety is the reason why the original three-form components $A_{\nu mn}$ have to be dualized in order for $d=4$ $\cN=8$ supergravity to exhibit a global $Sl(8)$-invariance \cite{CJ79,CJLP97}.\\

We want to make use of this $Sl(8)$-covariance for the comparison of the equations of motion of the truncation of $D=11$ supergravity to the version with manifest $E_{7(7)}$-covariance. In particular, this guarantees that it is sufficient to verify either the coefficient of the $F_{\alpha\beta cd}$-coupling or the one of the $F_{\alpha\beta}{}^a$-coupling in both the Einstein equation and the scalar equation of motion. We can hence without loss of generality focus on the $F_{\alpha\beta cd}$-coupling. The relevant terms in the Einstein equation (in the vielbein frame) immediately follows from the action $S$ (\ref{Actionneu}):
\beg
0
&=&
\frac14\left(\frac12 \eta^{\alpha\beta}R
-R^{\alpha\beta}\right)
+
\frac{1}{2\Delta}\left(F^\alpha{}_{\delta cd}F^{\beta\delta cd} -\frac14\eta^{\alpha\beta}F_{\g\delta cd}F^{\g\delta cd}\right)\\
&&+\text{further terms}.
\eeg
This coupling is then reproduced by a substitution of the twisted self-dual equation of motion (\ref{ImRe}) together with the identification (\ref{vSym26d}) in the $E_{7(7)}$-covariant form of the Einstein equation (\ref{BWGL4}). Hence, the Einstein equations of both theories agree. Since this coupling can be used to normalize the vector part $S_{\text{vec}}$ (\ref{action}) in the action $S_{\text{bos}}$ (\ref{action2}), a non-trivial statement is only obtained if the coupling in the scalar equations of motion coincide, too. This is indeed the case. To show this, we first link the Maurer--Cartan form of the scalars $(\cvA_\alpha)_{ABCD}:=e_\alpha{}^\mu (\cvA_\mu)_{ABCD}$ (\ref{BA}) to parts of the spin connection $\omega$ and the four-form field strength $F$ of $D=11$ supergravity by
\be\label{Fa2}
(\cvA_\alpha)_{ABCD}
&:=&
-\frac34\Delta^{-\frac12}\omega_{ef\alpha}
{\G^e}_{[AB}{\G^f}_{CD]}
\\
 &&
+\frac{1}{4}
\Delta^{-\frac12}F_{\alpha a_1\dots a_3}
{\G^{[a_1a_2}}_{[AB}{\G^{a_3]}}_{CD]}
\nn\\
&&
+\frac{i}{2880}
\Delta^{-\frac12}\left(-\frac{1}{3!}\e_{\alpha a_1\dots a_6 \beta_1\dots \beta_3 c}F^{\beta_1\dots \beta_3 c}\right)\e^{a_1\dots a_6c}
{\G_{bc}}_{[AB}{\G^{b}}_{CD]}.
\nn\ee
Then, we verify that this normalization of $\cvA$ reproduces the numerical factor in the coupling of the scalars to the Ricci tensor within the Einstein equation (\ref{BWGL4}), if we start from the action $S$ (\ref{Actionneu}) of $D=11$ supergravity. As a next step, it is straightforward to check that the coupling of the vectors to the scalars arises from the three-form equation of motion of $D=11$ supergravity (that is also derived from $S$ (\ref{Actionneu}))
\be\label{Maxwell}
\nabla_{B_1}F^{B_1\dots B_4}&=& -\frac{1}{24^2}\e^{B_2\dots B_4 A_1\dots A_8}F_{A_1\dots A_4}F_{A_5\dots A_8}.
\ee
Finally, we observe that this coupling of the vectors to the scalars agrees with the scalar equation of motion (\ref{selfdualC2c}) upon a substitution of the twisted self-dual equation of motion (\ref{ImRe}) together with the identification (\ref{vSym26d}). This proves that the equations of motion of the Kaluza--Klein reduction of $D=11$ supergravity on a flat seven-torus completely agree with the ones of the $E_{7(7)}$-covariant theory for the bosonic sector. Furthermore note that the coefficient linking the l.h.s. to the r.h.s. in the 3-form equation of motion (\ref{Maxwell}) is directly derived from the constant of the Chern--Simons term within the $D=11$ action $S$ (\ref{Actionneu}). This is another proof of the well-known statement that the global $E_{7(7)}$-symmetry of maximal supergravity would be absent for a different choice of Chern--Simons term coupling, as we have already mentioned at the end of section \ref{vielbeinS}.\\

Before addressing the fermions, we want to remark that the other two $Gl(7)$-representations $\cA_{\alpha,a}$ and $\cA_\alpha{}^{cd}$ within the $56$ vector potentials $\cA_\alpha{}^{AB}$ (\ref{ADec}) correspond to the so-called dual potentials. Given a field configuration of $D=11$ supergravity with $28$ vector potentials, the twisted self-duality equation of motion (\ref{ImRe}) then enables us to determine these dual potentials as ``non-local'' expressions of the $70$ scalars $G_{\cMa\cNa}$, the space-time metric $g_{\mu\nu}$ and the given $28$ vector fields. In view of the Hamiltonian analysis presented in this article, it is important to point out that we only have to relate the spatial components $\cA_i$ of the dual vector potentials to the given field configuration, because the zero component $\cA_0$ is not part of the action $S$. One can verify that these have the nice property to be expressible as spatial integrals on a constant time-slice only. For the particularly simple case of a configuration with constant almost complex structure $J^\cMa{}_\cNa$ of canoncial form $J=\binom{\,\,0\,\,\,\,1}{-1\,\,0}$ and vanishing shift $N^i$ (\ref{metric}), the twisted self-dual equation of motion (\ref{selfdualC}) leads e.g. to the relation
\be\label{selfdualC24}
\cF_{ij}{}^{m_1m_2} &=&-\frac{1}{e_4}h_{ij_1}h_{jj_2}\e^{j_1j_2k}\eta^{m_1n_1}\eta^{m_2n_2} \cF_{0k,m_1m_2}.
\ee
Since this form is closed $\p_{[k}\cF_{ij]}{}^{m_1m_2}=0$ by the equations of motion, the dual potential $\cA_{j}{}^{m_1m_2}$ in the curved frame (\ref{v114}) would for this case be uniquely defined by this equation (\ref{selfdualC24}) as a spatial integral on a constant time-slice of the metric $g_{\mu\nu}$ and the usual $21$ four-form field strengths $F_{\alpha\beta cd}$ (\ref{vSym26d}).\\

The statement of equivalent dynamics of the reduction of $D=11$ supergravity and our $E_{7(7)}$-invariant theory extends to the fermionic sector. To show this, we relate the supersymmetry parameter $\ep$ and the gravitino $\psi_M$ of $D=11$ supergravity to the fermions $\eps,\chi$ used in the action $S_{\text{ferm}}$ (\ref{actionferm}) in the standard way \cite{CJ79,dWN86}:
\be\label{rescale4}
{\eps}^A &:=& \frac{1}{2}\sqrt{-\g_5}\Delta^{\frac{1}{4}}\left(\id_4 -i\g_5\right){\ep}^A \\
(\chi_a)^A&:=&\frac{1}{2}\sqrt{-\g_5}\Delta^{-\frac{1}{4}}\left(\id_4 -i\g_5\right)(\psi_a)^A
\nn\\
(\chi_\alpha)^A&:=&\frac{1}{2}\sqrt{-\g_5}\Delta^{-\frac{1}{4}}\left(\id_4 -i\g_5\right)\left((\psi_\alpha)^A +\frac{i}{2}\g_5\g_\alpha {{\G^a}^A}_B(\psi_a)^B\right)
\nn\\
\text{with}\quad \sqrt{-\g_5}&:=& \frac{1}{\sqrt{2}}\left(\id_4 -\g_5\right)
\nn
\ee
and with the vector indices ranging over $\alpha=0,\dots,3$ and $a=4,\dots,10$ and the spinor indices $A,B=1,\dots,8$ that arise from writing the $32$ dimensional spinors $\ep,\psi_M$ in $D=11$ as $8$ four-dimensional spinors. The $7\times 8$ fermions $(\chi_a)^C$ are then combined into an $SU(8)$-representation by
\be\label{sumFermi}
\chi^{ABC} &:=& 3!i{\G^a}^{[AB}(\chi_a)^{C]}.
\ee
The Clifford algebra representation matrices $\tilde{\G}\in \R^{32\times 32}$ in eleven dimensions can be decomposed into the ones of $d=4$ $\g^\alpha\in \R^{4\times 4}$ and the $d=7$ ones $\G^a\in i\R^{8\times 8}$ in the standard way
\begin{subequations}\label{gammadec2}
\be
\tilde{\G}^\alpha &=& \g^\alpha\otimes \id_{8}\quad \,\text{for }\alpha\,=\,0,\dots,3
\\
\tilde{\G}^a &=& \frac{\g_5}{i}\otimes \G^a\quad \text{for }a\,=\,4,\dots,10.
\ee
\end{subequations}
Here, we should keep in mind that the $8\times 8$ matrices $\G^a$ are purely imaginary and $\g_5^2=-\id_4$ (\ref{clifford}). The fact that we are using the Majorana spinor formalism for the $d=4$ spinors implies that $\ep^C,(\psi_a)$ are real quantities. Following the convention introduced for the scalar sector of the bosons in (\ref{vielbeinG}), a complex conjugation changes the position of the $SU(8)$-indices $A,B,C$. Together with the definition (\ref{rescale4}), this leads e.g. to the identity $\eps_A=\frac{1}{2}\sqrt{-\g_5}\Delta^{\frac{1}{4}}\left(\id_4 +i\g_5\right){\ep}_A$ where the position of the index $A$ for the real Majorana spinor $\ep$ is arbitrary, of course. Furthermore, the definition (\ref{rescale4}) immediately satisfies the identities $P^+\eps^A=0$, $P^+(\chi_\alpha)^A=0$ and $P^+\chi^{ABC}=0$ with the projector $P^+:=\frac12(\id_4+i\g_5)$.\\

The reader may have noticed that we are using the same notation $P^\pm$ for the projectors $P^\pm=\frac12(\id_4\pm i\g_5)$ and for $(P^\pm)_{ \beta_1\beta_2}^{\beta_3 \beta_4}=\frac12 (
\delta_{ \beta_1\beta_2}^{\beta_3 \beta_4}
\pm\frac{i}{2}\e_{ \beta_1\beta_2}{}^{\beta_3 \beta_4}
	 	  	  	  )$ (\ref{projector}). This should not come as a surprise due to the following identity for $\g$-matrices in $d=4$:
	 	  	  	  \beg
	 	  	  	  \g_5\g_{\beta_1\beta_2}&=&\frac{1}{2}\e_{ \beta_1\beta_2}{}^{\beta_3 \beta_4}\g_{\beta_3\beta_4}.
	 	  	  	  \eeg
	 	  	  	  For the proof of the agreement of the bifermionic coupling to the vector fields that is provided by $W_{\beta_1\beta_2}{}^{AB}$ (\ref{WpDefi2}) in our case and by $O^+$ (2.22) in \cite{dWN82}, the following identities for ``holomorphic'' or ``chiral'' spinors $\chi$, i.e. for the ones with raised $SU(8)$ index, have been useful:
\begin{subequations}\label{Proj2}
	 	  	  	  \be
	 	  	  	  (P^+)_{ \beta_1\beta_2}^{\beta_3 \beta_4}\g_{\beta_3 \beta_4}\chi &=& 0\\
	 	  	  	  (P^-)_{ \beta_1\beta_2}^{\beta_3 \beta_4}\g_{\beta_3 \beta_4}\chi &=& \g_{\beta_1 \beta_2}\chi\\
	 	  	  	  \label{proj3}
	(P^+)^{ \beta_1\beta_2}_{\beta_3 \beta_4}\g^{\beta_3 \beta_4}\g^{\beta_5}\chi&=& 4(P^+)^{ \beta_1\beta_2}_{\beta_3 \beta_4}\g^{\beta_3}\eta^{ \beta_4\beta_5}\chi.
	 	  	  	  \ee
	 	  	  	  \end{subequations}
	 	  	  	  
	 	  	  	  In the same way as the equations of motion, the supersymmetry variations (\ref{SFerm3},\,\ref{c23}) can be obtained from the ones (\ref{Trafo1neu}) of $D=11$ supergravity by a Kaluza--Klein reduction  on $T^7$. There is however a subtlety to keep in mind concerning this procedure. It is crucial to impose a block-diagonal matrix form for the elfbein $E$ (\ref{v114b}) by fixing the local Lorentz symmetry $SO(10,1)$ to $SO(3,1)\times SO(7)$ \cite{CJ79,dWN86}. This in particular implies $E_m{}^\alpha=0$ for $\alpha=0,\dots,3$ and $m=4,\dots,10$. In order for this to be consistent with the supersymmetry variation $\delta^{(11)}$ of the elfbein $E$, one has to modify the definition of the supersymmetry variation in such a way that $\delta^{(4)}E_m{}^\alpha=0$. This is obtained by relating the two supersymmetry variations $\delta^{(11)}$ and $\delta^{(4)}$ by a compensating $\mathfrak{so}_{(10,1)}$-rotation $\Sigma$ with the only non-vanishing components $\Sigma^{\alpha b}:=E^{mb}\delta^{(11)}E_m{}^\alpha=-\Sigma^{b \alpha}$ \cite{dWN86}:
\be\label{susy4}
\delta^{(4)}E_M{}^A&:=& \delta^{(11)}E_M{}^A -\Sigma^{AB} E_{MB}\\
\text{with}\quad M,A,B&=&0,\dots,10.
\nn
\ee
The compensating $\mathfrak{so}_{(10,1)}$-rotation $\Sigma^{AB}$ is linear in fermions $\chi$ (\ref{rescale4}), but obviously not $SU(8)$ covariant. The modification of the supersymmetry variation in passing from $D=11$ to $d=4$ is hence completely analogous to the introduction of the covariant supersymmetry variation $\deltaS$ in eq. (\ref{SDelta}). 
\end{section}


\begin{thebibliography}{21}
\bibitem{CJ79}  E.~Cremmer and B.~Julia,
  {\sl The $SO(8)$ supergravity},\\ Nucl. Phys. B {\bf 159} (1979) 141

\bibitem{dWN82}  B.~de Wit and H.~Nicolai,
  {\sl $\cN=8$ supergravity},\\ Nucl. Phys. B {\bf 208} (1982) 323

\bibitem{CJS78} E.~Cremmer, B.~Julia, J.~Scherk, {\itshape Supergravity Theory in 11 dimensions}, Phys. Lett. {\bf B76} (1978) 409
  
\bibitem{CJLP97} E.~Cremmer, B.~Julia, H.~Lu and C.~N.~Pope,
  {\sl Dualisation of dualities. I},
  Nucl.\ Phys.\  B {\bf 523} (1998) 73,
  [arXiv:hep-th/9710119]

\bibitem{HT88}
  M.~Henneaux and C.~Teitelboim,
  {\sl Dynamics of chiral (selfdual) p forms},
  Phys. Lett. B {\bf 206} (1988) 650
  
\bibitem{ADM62}  R.~Arnowitt, S.~Deser and C.~W.~Misner,
  {\sl The Dynamics of General Relativity}, in: L. Witten (ed.), Gravitation (Wiley, NY, 1962) 227
  
\bibitem{H08}
  C.~Hillmann,
  {\sl $E_{(7(7))}$ and $D=11$ supergravity},
  PhD thesis,\\ Humboldt University of Berlin,
  [arXiv:0902.1509]

\bibitem{CJLP98} E.~Cremmer, B.~Julia, H.~Lu and C.~N.~Pope,
  {\sl Dualisation of dualities II: Twisted self-duality of doubled fields  and
  superdualities},\\
  Nucl.\ Phys.\  B {\bf 535} (1998) 242,
  [arXiv:hep-th/9806106]
    
\bibitem{dWN86}  B.~de Wit and H.~Nicolai,
  {\sl $d=11$ supergravity with local $SU(8)$ invariance}, Nucl. Phys. B {\bf 274} (1986) 363

\bibitem{dW02}  B.~de Wit,
  {\sl Supergravity}, Les Houches Lecture notes 2001,\\ {}[arXiv:~hep-th/0212245]
  
\bibitem{GZ81}
  M.~K.~Gaillard and B.~Zumino,
  {\sl Duality Rotations For Interacting Fields},
  Nucl.\ Phys.\  B {\bf 193} (1981) 221

\bibitem{KS08}
  R.~Kallosh and M.~Soroush,
  {\sl Explicit action of $E_{7(7)}$ on $\cN=8$ supergravity fields},
  Nucl.\ Phys.\  B {\bf 801} (2008) 25
  [arXiv:0802.4106]
  
\bibitem{D64} P.~A.~M.~Dirac, {\sl
    Lectures on Quantum Mechanics},\\ Belfer Graduate School of Science (1964)  
  
\bibitem{MN94}
  H.~J.~Matschull and H.~Nicolai,
  {\sl Canonical treatment of coset space sigma models},
  Int.\ J.\ Mod.\ Phys.\  D {\bf 3} (1994) 81
  
\bibitem{T01}
  T.~Thiemann,
  {\sl Introduction to modern canonical quantum general relativity},
  [arXiv:gr-qc/0110034]

\bibitem{PSS96}
J.~M.~Pons, D.~C.~Salisbury and L.~C.~Shepley,
{\sl Gauge transformations in the Lagrangian and Hamiltonian formalisms of
generally covariant theories},
Phys.\ Rev.\  D {\bf 55} (1997) 658,
[arXiv:gr-qc/9612037]

\bibitem{PSS99}
J.~M.~Pons, D.~C.~Salisbury and L.~C.~Shepley,\\
{\sl Gauge transformations in Einstein-Yang-Mills theories},\\
J.\ Math.\ Phys.\  {\bf 41} (2000) 5557,
[arXiv:gr-qc/9912086]


\bibitem{HT87}
 M.~Henneaux and C.~Teitelboim,
  {\sl Consistent Quantum Mechanics Of Chiral P Forms},
  In: *Santiago 1987, Proceedings, Quantum mechanics of fundamental systems 2* 79-112.
  
\bibitem{Finite}
   Z.~Bern, J.~J.~Carrasco, L.~J.~Dixon, H.~Johansson, D.~A.~Kosower and R.~Roiban,
  {\sl Three-Loop Superfiniteness of $\cN=8$ Supergravity},\\
  Phys. Rev. Lett. {\bf 98} (2007) 161303,
  [arXiv:hep-th/0702112 ]
;\\
  M.~B.~Green, J.~G.~Russo and P.~Vanhove,
  {\sl Ultraviolet properties of maximal supergravity},
  Phys. Rev. Lett.  {\bf 98} (2007) 131602  
  
\bibitem{FiniteE7}
   N.~Arkani-Hamed, F.~Cachazo and J.~Kaplan,
  {\sl What is the Simplest Quantum Field Theory?},
  [arXiv:0808.1446];\\
  R.~Kallosh, C.~H.~Lee and T.~Rube, 
  {\sl $\cN=8$ Supergravity 4-point Amplitudes}, [arXiv:0811.3417]
  
\bibitem{KK08}
R.~Kallosh and T.~Kugo,
  {\sl The footprint of $E_{7(7)}$ in amplitudes of $\cN=8$ supergravity},
    JHEP {\bf 0901} (2009) 072,
  [arXiv:0811.3414]    
 
\bibitem{K81}
R.~Kallosh,
  {\sl Counterterms in extended supergravities},\\
    Phys.Lett. B {\bf 99},2 (1981) 122 
 
\bibitem{BKR08} 
  L.~Brink, S.~S.~Kim and P.~Ramond,
  {\sl $E_{7(7)}$ on the Light Cone},\\
  JHEP {\bf 0806} (2008) 034,
    [arXiv:0801.2993]
    
\bibitem{dWST02}  
B.~de Wit, H.~Samtleben and M.~Trigiante,\\
  {\sl On Lagrangians and gaugings of maximal supergravities},\\
  Nucl.\ Phys.\  B {\bf 655} (2003) 93,
  [arXiv:hep-th/0212239] 

\bibitem{dWST07} 
 B.~de Wit, H.~Samtleben and M.~Trigiante,\\
  {\sl The maximal D=4 supergravities},\\
  JHEP {\bf 0706} (2007) 049,
  [arXiv:0705.2101]
    
\bibitem{DHN02} T.~Damour, M.~Henneaux and H.~Nicolai, {\sl
    $E_{10}$ and a ``small tension expansion'' of M-theory}, Phys.
    Rev. Lett. {\bf 89} (2002) 221601, [ hep-th/0207267]
  
\bibitem{W01}  P.~C.~West,
  {\sl $E_{11}$ and M theory}, Class. Quant. Grav. {\bf 18} (2001) 4443
    
\bibitem{H09}
  C.~Hillmann,
  {\sl Generalized E(7(7)) coset dynamics and D=11 supergravity},
  JHEP {\bf 0903} (2009) 135,
  [arXiv:0901.1581]
 
\bibitem{FN76} 
D.~Z.~Freedman, P.~van Nieuwenhuizen and S.~Ferrara,
  {\sl Progress Toward A Theory Of Supergravity},
  Phys. Rev.  D {\bf 13} (1976) 3214;\\
D.Z. Freedman, P.~van Nieuwenhuizen, {\itshape Properties of supergravity theory}, Phys. Rev. D {\bf 14} (1976) 912
\end{thebibliography}
\end{document}